\documentclass[twocolumn]{aastex631} 
\usepackage{amsmath}
\usepackage{amssymb}
\usepackage{bm}
\usepackage{booktabs}
\usepackage[caption=false]{subfig}
\usepackage{float}
\usepackage{tabularx}

\newcommand{\rion}[2]{{\ensuremath{\mbox{\rm #1$\,${\small\uppercase\expandafter{\romannumeral#2\relax}}}}}}

\newcommand{\HII}{\rion{H}{2}}

\newcommand{\HeII}{\rion{He}{2}}
\newcommand{\CIV}{\rion{C}{4}}
\newcommand{\CIII}{\rion{C}{3}}

\newcommand{\OIII}{[\rion{O}{3}]}
\newcommand{\OII}{[\rion{O}{2}]}

\newcommand{\SII}{[\rion{S}{2}]}

\newcommand{\NeIII}{[\rion{Ne}{3}]}

\newcommand{\Msun}{${\rm M}_{\odot}$}

\newcommand{\BEAGLE} {\ensuremath{\mbox{\rm \texttt{BEAGLE}}}}

\shortauthors{Ch\'{a}vez Ortiz et al.}
\begin{document}

\title{Significant Evidence of an AGN Contribution in GHZ2 at $z =$ 12.34}

\correspondingauthor{\'{O}scar A. Ch\'{a}vez Ortiz}
\email{chavezoscar009@utexas.edu}

\author[0000-0003-2332-5505]{\'{O}scar A. Ch\'{a}vez Ortiz}
\altaffiliation{NASA FINESST Fellow}
\affiliation{Department of Astronomy, The University of Texas at Austin, 2515 Speedway Boulevard, Austin, TX 78712, USA}
\affiliation{Cosmic Frontier Center, The University of Texas at Austin, Austin, TX 78712, USA}

\author[0000-0001-8519-1130]{Steven L. Finkelstein}
\affiliation{Department of Astronomy, The University of Texas at Austin, 2515 Speedway Boulevard, Austin, TX 78712, USA}
\affiliation{Cosmic Frontier Center, The University of Texas at Austin, Austin, TX 78712, USA}

\author[0000-0003-0390-0656]{Adele Plat}
\affiliation{Institute of Physics, GalSpec Laboratory, Ecole Polytechnique Federale de Lausanne, Observatoire de Sauverny, Chemin Pegasi 51, 1290 Versoix, Switzerland}

\author[0009-0002-0651-5761]{Maddie Silcock}
\affiliation{Centre for Astrophysics Research, Department of Physics, Astronomy and Mathematics, University of Hertfordshire, Hatfield AL10 9AB, UK}

\author[0000-0002-9551-0534]{Emma Curtis Lake}
\affiliation{Centre for Astrophysics Research, Department of Physics, Astronomy and Mathematics, University of Hertfordshire, Hatfield AL10 9AB, UK}

\author[0000-0003-1282-7454]{Anthony Taylor}
\affiliation{Department of Astronomy, The University of Texas at Austin, 2515 Speedway Boulevard, Austin, TX 78712, USA}

\author[0000-0003-4242-8606]{Ansh R. Gupta}
\affiliation{Department of Astronomy, The University of Texas at Austin, 2515 Speedway Boulevard, Austin, TX 78712, USA}
\affiliation{Cosmic Frontier Center, The University of Texas at Austin, Austin, TX 78712, USA}

\author[0000-0002-8951-4408]{Lorenzo Napolitano}
\affiliation{INAF – Osservatorio Astronomico di Roma, via Frascati 33, 00078, Monteporzio Catone, Italy}
\affiliation{Dipartimento di Fisica, Università di Roma Sapienza, Città Universitaria di Roma - Sapienza, Piazzale Aldo Moro, 2, 00185, Roma, Italy}

\author[0000-0001-9875-8263]{Marco Castellano}
\affiliation{INAF - Osservatorio Astronomico di Roma, via di Frascati 33, 00078 Monte Porzio Catone, Italy}

\author[0000-0003-0212-2979]{Volker Bromm}
\affiliation{Department of Astronomy, The University of Texas at Austin, 2515 Speedway Boulevard, Austin, TX 78712, USA}
\affiliation{Cosmic Frontier Center, The University of Texas at Austin, Austin, TX 78712, USA}
\affiliation{Weinberg Institute for Theoretical Physics, University of Texas, Austin, TX 78712, USA}

\author[0000-0001-7300-9450]{Ikki Mitsuhashi}
\affiliation{Department for Astrophysical \& Planetary Science, University of Colorado, Boulder, CO 80309, USA}

\author[0000-0003-3458-2275]{Stephane Charlot}
\affiliation{Sorbonne Universit\'e, CNRS, UMR 7095, Institut d'Astrophysique de Paris, 98 bis bd Arago, 75014 Paris, France}

\author[0000-0003-3820-2823]{Adriano Fontana}
\affiliation{INAF – Osservatorio Astronomico di Roma, via Frascati 33, 00078, Monteporzio Catone, Italy}

\author[0000-0002-7051-1100]{Jorge A. Zavala}
\affiliation{University of Massachusetts Amherst, 710 North Pleasant Street, Amherst, MA 01003-9305, USA}
\email{jzavala@umass.edu}  

\author[0000-0002-7636-0534]{Jacopo Chevallard}
\affiliation{Department of Physics, University of Oxford, Denys Wilkinson Building, Keble Road, Oxford OX1 3RH, UK}

\author[0000-0002-4193-2539]{Denis Burgarella}
\affiliation{Laboratoire d’Astrophysique de Marseille, Aix Marseille University, CNRS, CNES, 38, rue Fred´ eric ´ Joliot-Curie, F-13388 Marseille, cedex 13, France}

\author[0000-0002-3301-3321]{Michaela Hirschmann}
\affiliation{Institute of Physics, Laboratory for Galaxy Evolution, EPFL, Observatoire de Sauverny, Chemin Pegasi 51, CH-1290 Versoix, Switzerland}

\author[0000-0002-5268-2221]{Tom Bakx}
\affiliation{Department of Space, Earth and Environment, Chalmers University of Technology, SE-412 96 Gothenburg, Sweden}

\author[0000-0002-5268-2221]{Alba Vidal-Garc\'ia }
\affiliation{Observatorio Astronómico Nacional, C/ Alfonso XII 3, 28014 Madrid, Spain}

\author[0000-0003-2536-1614]{Antonello Calabr\`o}
\email{antonello.calabro@inaf.it}
\affiliation{INAF Osservatorio Astronomico di Roma, Via Frascati 33, 00078 Monte Porzio Catone, Rome, Italy}

\author[0000-0001-6865-2871]{Anna Feltre}
\affiliation{INAF—Osservatorio Astrofisico di Arcetri, Largo E. Fermi 5, I-50125, Firenze, Italy}

\begin{abstract}
\noindent GHZ2 is among the highest-redshift galaxies discovered to date, exhibiting a spectrum rich with prominent emission lines in the rest-frame ultraviolet (UV) and optical. These features raise critical questions about the mechanism powering this nebular emission, in particular the extremely strong \ion{C}{4}$\lambda$1548+1551 emission (rest-frame EW $=$ 45 \AA). Here we  aim to quantify the AGN contribution within this system using the \BEAGLE-AGN tool to simultaneously fit the spectrum and photometry of GHZ2.  We consider a range of models with and without AGN components, allowing us to disentangle the stellar and AGN contribution of GHZ2 for the first time  simultaneously using all observed emission lines. We conclude that a partial contribution by an AGN is significantly favored based on 
$\chi^2$
comparisons between models with and without an AGN component, measuring an AGN contribution of   97$_{-2}^{+1}$ \%, and 64$_{-12}^{+23}$ \% for the \CIV$\lambda$1548+1551 and \CIII]$\lambda$1908 emission lines, respectively. 
 We estimate the black hole mass using the accretion luminosity ($L_{acc}$) from the best fit \BEAGLE-AGN model, 
 inferring a value of log$_{10}$(M$_{BH}$/M$_{\odot}$) = 7.19$^{+0.03}_{-0.03}$, assuming an Eddington ratio of $\eta$ = 0.5 (with a  much larger systematic uncertainty of $\sim$1 dex). The inferred black hole mass to stellar mass ratio is 0.06$^{+0.02}_{-0.02}$ ( statistically), consistent with other high redshift AGN systems. If the black hole interpretation is confirmed, GHZ2 would represent the most distant black hole identified to date, making it an ideal laboratory to study AGN growth and their role in shaping high-redshift galactic evolution.
\end{abstract}


\section{Introduction}

Since the launch of {\it JWST}, the frontier of what is measurable for galactic evolution in the early universe has undergone a paradigm shift. {\it JWST} is finding evidence that accreting black holes (BH) exist at earlier times and at masses higher than what was expected, given the lack of time that has elapsed at high redshift. Spectroscopic follow-up of high redshift galaxies has also shown that many of these systems host an active galactic nuclei (AGN), as evidenced through the presence of broad Balmer emission \citep[e.g.][]{Kokorev_2023, Larson_AGN_2023, Maiolino_2024,Fujimoto_2024, Taylor_A_2024, taylor_lrdz9_2025}, implying that high redshift black holes may be playing a larger role in the evolution of galaxies than previously thought. Studying systems that harbor an AGN requires careful analysis and modeling to decouple the AGN contribution from the stellar contribution. These `composite' systems often reside in a region of parameter space where star forming (SF) galaxies and AGN overlap in diagnostic line-ratio diagrams, which makes determining the main ionization mechanism difficult, notable examples include UHZ1 and GHZ9 \citep[][]{Bogdan_2024, Napolitano_2024, Kovacs_2024}. 

A complicating factor when using locally calibrated line ratios to determine the ionization source at high redshift is that the diagnostics that effectively separate AGN and star-formation lose their classification power at higher redshift.  For example, low metallicity AGN can fall in the SF region of common line ratio diagrams \citep[e.g.,][]{Backhaus_OHNO_2022, Cleri_2025}. A compounding challenge is that the emission lines which most-accurately differentiate between AGN and star formation are often detected only at low significance, or are out of the spectral range of the instruments. Thus, traditional line ratio diagnostics struggle to discern whether the radiation is dominated by stars or AGN in these composite systems \citep[see][]{Ubler_2023, Castellano_2024, Scholtz_2025}. However, tools are now being developed which can model the full spectra of galaxies, constraining both an AGN and stellar component, such as \BEAGLE-AGN \citep[][]{vidal_garcia_2024, Silcock_2025} and \texttt{CIGALE} \citep{Boquien_cigale_2019, Yang_cigale_2022}.

An exceptional case study to test these tools is a galaxy at redshift 12.34, GLASS-z12 (GHZ2), discovered through data collected as part of the GLASS survey \citep[][]{Castellano_2022, Naidu_2022,Treu_2022, Castellano_2024}. One of the key features that makes this galaxy unique is that it was the first $z >$ 10 galaxy with oxygen emission lines detected using MIRI spectroscopy \citep{Zavala_2024}. In addition, GHZ2 shows prominent high-ionization UV lines such as \CIV$\lambda$1548+1551 (EW = 45.8 \AA, $>$47.9~eV) 
in its NIRSpec spectrum, indicative of a very strong ionization source powering the UV lines (\citealt{Castellano_2024}). This naturally raises the question: what is powering the ionizing radiation that produces the nebular emission lines in the spectrum? Is the radiation powered by a stellar contribution alone, or by an AGN? \citet{Castellano_2024} performed a careful line-ratio analysis using the \CIV$\lambda$1548+1551, \CIII]$\lambda$1908, and \HeII$\lambda$1640 emission lines and found that this source is in a region of parameter space where the photoionization models of star formation and AGN models overlap. 
A complementary analysis using the NIRSpec and MIRI data by \citet[][]{Calabro_2024} found a similar result to \citet[][]{Castellano_2024}, where GHZ2 lies in a region of parameter space in the ``OHNO" diagram \citep[][]{Backhaus_OHNO_2022} explained by AGN and SF photoionization models. Finally, \citet{Zavala_OIIImicrons_2024} inferred from the ALMA-detected \OIII$\lambda$88$\mu$m emission line the ionization source is consistent with compact star star-formation, though they noted that their observations did not rule out some AGN activity. Thus, the true nature of this extreme source is still unknown, and more advanced modeling tools are needed.

These previous studies used line ratios to discern the ionization mechanism within GHZ2.  This paper aims to see if we can move beyond typical line-ratio diagrams by simultaneously modeling the  continuum and emission lines in the NIRSpec spectrum and  using line fluxes derived from the MIRI spectrum to constrain the rest-optical.  This joint modeling will enable more robust constraints on whether an AGN is required to explain the nebular emission of GHZ2. 

Motivated by the strong \CIV$\lambda$1548+1551 emission, we aim to quantify the amount of AGN contribution, if any, required to explain this line which is also consistent with all other measured line fluxes. We use the \BEAGLE-AGN tool that can model the full spectrum accounting for both AGN and star formation-powered ionization \citep[][]{vidal_garcia_2024, Silcock_2025}.

The structure of this paper is as follows: Section~\ref{sec:data} outlines the photometric and spectroscopic data used in this analysis. Section~\ref{sec:methods} describes the \BEAGLE-AGN modeling and the analysis to quantify the AGN contribution. Section~\ref{sec:results} presents the results of this analysis and Section~\ref{sec:discussion} discusses the broader implications of our finding and next steps for understanding high redshift galaxies. 

We assume a cosmological model with $H_0$ = 70 km s$^{-1}$~Mpc$^{-1}$, $\Omega_{m, 0}$ = 0.3, and $\Omega_{\Lambda, 0}$ = 0.7, and all reported magnitudes are expressed in the AB magnitude system \citep{OKE_1974, Oke_Gunn_1983}. 

\section{Data} \label{sec:data}

Studying the impact of a potential AGN within GHZ2 requires spectroscopic and photometric data that covers the rest-frame UV and optical wavelengths. This combination of data enables us to perform spectro-photometric fitting with \BEAGLE-AGN (\S 3.1) to constrain the properties of the stellar and AGN components.

\subsection{NIRSpec Spectroscopy}
The NIRSpec PRISM spectroscopic data for GHZ2 were taken by PI: \citet[][]{Castellano_2024} as part of GO-program 3073, which obtained two overlapping NIRSpec PRISM pointings on the GLASS-{\it JWST} NIRCam field, with observations divided into three visits per pointing to enable six-band parallel NIRCam imaging. Seven usable dithers of galaxy GHZ2 were analyzed, totaling 15,323 seconds of exposure time, despite one visit being partially compromised by an electrical short. Data were processed using the STScI Calibration Pipeline, including corrections for detector noise, flat-fielding, and background subtraction. A low-redshift contaminant (z = 1.68) required masking and custom background subtraction to mitigate spurious features. The final spectra were matched to the NIRCam photometry to correct for wavelength-dependent slit and aperture losses, with corrections ranging from 1.19 at 2 $\mu$m to 1.78 at 5 $\mu$m. These steps produced clean 2D and 1D spectra for analysis as described in more detail by \cite{Castellano_2024}. The NIRSpec wavelength range spans rest-frame 489 - 3968 \AA, fully covering the UV and partially covering the optical.  Since we are motivated by the high ionization rest frame UV emission lines, we use the NIRSpec spectrum in the spectro-photometric fitting with \BEAGLE-AGN to constrain the UV, while MIRI is used to constrain the optical.  The NIRSpec spectrum used in this analysis has been corrected to the photometry by performing a wavelength dependent correction, ensuring consistency between the photometry and the NIRSpec spectrum.

Within the NIRSpec spectrum there are significant ($>$ 3$\sigma$ detected) emission lines, such as: \ion{N}{4}]$\lambda$1486, \CIV$\lambda$1548+1551, \HeII$\lambda$1640, \ion{O}{3}]$\lambda$1660+1666, \ion{N}{3}]$\lambda$1750, \CIII]$\lambda$1908, \ion{O}{3}$\lambda$3133, \OII$\lambda$3727+3729, \NeIII$\lambda$3869.  While \citet{Castellano_2024} fit these lines, to ensure consistency between the NIRSpec and MIRI flux measurements (Section \ref{sec:miri_spec}) in this analysis, we remeasure the line fluxes using \texttt{emcee} \citep[][]{emcee_2013}. Due to the low resolution of the NIRSpec, intrinsic doublets such as the \ion{O}{3}]$\lambda$1660+1666 lines are modeled as a single Gaussian. All the lines have been fit with a single Gaussian, except for the \ion{He}{2}$\lambda$1640+\ion{O}{3}]$\lambda$1660+1666 where a joint fit using two Gaussians is used, one Gaussian to cover the \ion{He}{2}$\lambda$1640 and another Gaussian to cover the \ion{O}{3}]$\lambda$1660+1666 doublet. We find line fluxes consistent with previous measurements in the literature \citep[][]{Castellano_2024}.

Prior to fitting with \BEAGLE-AGN, we mask out data blue-ward of the Lyman-break and mask out the region around [\ion{Ne}{3}]$\lambda$3869 due to a known  issue within \BEAGLE-AGN where it systematically overestimates the metallicity in the narrow-line region and has a discrepancy between observed high-redshift [\ion{Ne}{3}]$\lambda$3869  fluxes and current \BEAGLE-AGN model coverage \citep[][]{Silcock_2025}.  An independent fit to quantify the impact of including the [\ion{Ne}{3}]$\lambda$3869 line was conducted. To verify that this choice does not influence our conclusions, we repeated the full BEAGLE-AGN analysis including [Ne III]$\lambda3869$. The inferred posterior distributions of the primary AGN and galaxy parameters changed negligibly, demonstrating that our scientific conclusions are not driven by the inclusion or exclusion of this line.

\begin{table}
\centering
\caption{Measured emission-line fluxes and their associated $1\sigma$ uncertainties. Fluxes are given in units of $10^{-18}\,\mathrm{erg\,s^{-1}\,cm^{-2}}$. The H$\beta$ and \OIII 52$\mu$m measurements are upper limits.}
\label{tab:obs_flux}
\begin{tabular}{lcc}
\hline
Line & Flux & $1\sigma$ Error \\
\hline
N\,IV] $\lambda1486$            & 0.77 & 0.26 \\
C\,IV $\lambda1548+1551$            & 2.59 & 0.17 \\
He\,II $\lambda1640$           & 0.34 & 0.32 \\
O\,III] $\lambda1660+1666$          & 0.65 & 0.20 \\
N\,III] $\lambda1750$          & 0.55 & 0.21 \\
C\,III] $\lambda1908$          & 0.85 & 0.14 \\
Mg\,II $\lambda2795+2802$           & 0.26 & 0.57 \\
$[$O\,II$]$ $\lambda3727+3729$      & 0.58 & 0.31 \\
H$\beta$                       & 0.00 & 0.76 \\
$[$O\,III$]$ $\lambda4959$     & 1.80 & 0.21 \\
$[$O\,III$]$ $\lambda5007$     & 5.41 & 0.62 \\
H$\alpha$                      & 2.59 & 0.71 \\
$[$O\,III$]$ $52\,\mu$m        & 0.00 & 0.57 \\
$[$O\,III$]$ $88\,\mu$m        & 0.30 & 0.07 \\
\hline
\end{tabular}
\end{table}

\subsection{MIRI Spectroscopy}\label{sec:miri_spec}

Observations for project GO-3703 (PI: J. Zavala) used the MIRI Low Resolution Spectrometer (P750L filter) in slit mode, with three visits totaling 9 hours of on-source time. Data reduction employed {\it JWST} pipeline version 1.13.4 and CRDS context \texttt{${\it JWST}\_1174.pmap$}. After standard Stage 1 and 2 reductions, the presence of a bright nearby galaxy in one dither required separate processing of each dither. Background subtraction was performed using \texttt{photutils}, and the Stage 3 pipeline combined the two dither positions into a single 2D spectrum. A manual 1D extraction with a boxcar filter adjusted to the MIRI PSF provided improved r.m.s. noise compared to the standard pipeline. Above 9 $\mu$m, where noise increased, data were rebinned into 2-channel bins. A pipeline-based reduction showed a slightly brighter [OIII]$\lambda$5007 line luminosity but lower signal-to-noise ratio. The manual reduction was adopted for analysis due to better overall noise performance. We refer the reader to \cite{Zavala_2024} for additional details about the data reduction. The full MIRI coverage of GHZ2 spans rest-frame 3796 -- 8990 \AA. 

 As \BEAGLE-AGN can not model the different resolution curves between the NIRSpec and MIRI spectroscopic data, we do not fit the full MIRI spectrum, but rather pass \BEAGLE-AGN the integrated emission lines for H$\beta$, \OIII$\lambda$5007 and  H$\alpha$, which it uses to constrain the rest-optical emission lines.  The optical line fluxes have been remeasured using \texttt{emcee} and we jointly fit H$\beta$, \OIII$\lambda$4960, and \OIII$\lambda$5007 together with three separate Gaussian functions and a linear continuum. We fix the intrinsic flux ratio of the \OIII$\lambda$5007/\OIII$\lambda$4960 to 2.98. We fit the H$\alpha$ line with a single Gaussian and a linear continuum. We measure line fluxes consistent with previously reported line measurements from \citep[][]{Zavala_2024}.
All fluxes reported are in units of $10^{-18}$ erg\,s$^{-1}$\,cm$^{-2}$  and are used in the fitting as constraints for the optical regime.  We input the line fluxes for \OIII$\lambda$5007 = 5.41$\pm$ 0.62, and H$\alpha$ = 2.59$\pm$ 0.71, as well the measured upper limit for H$\beta$ $\leq$ 0.76.  We note that there are also ALMA observations of this source by \citet{Zavala_OIIImicrons_2024} who found an upper limit on the \OIII 52$\mu$m and detected \OIII 88$\mu$m emission. We use the values from \citet{Zavala_OIIImicrons_2024} and fold them into the fitting for further constraining power, the final observed measurements are seen in Table~\ref{tab:obs_flux}.

\subsection{Photometric Data}

The GLASS-{\it JWST}-ERS \citep[][]{Treu_2022} program utilizes NIRCam's LW and SW channels with a set of seven wide filters (F090W, F115W, F150W, F200W, F277W, F356W, and F444W) for imaging in parallel with NIRSpec and NIRISS observations. For NIRCam imaging parallel to NIRSpec, six exposure slots (8,245 s each) result in total imaging times of $\approx$4.6 hr for F090W and F115W, $\approx $2.3 hr for F150W, F200W, and F356W, and $\approx$ 9.2 hr for F444W, achieving 5$\sigma$ AB magnitude limits of $\approx$ 29.2–29.7 for point sources. Parallel to NIRISS, imaging includes six groups per exposure, totaling $\approx$ 6.5 hr for F444W, $\approx$ 3.2 hr for F090W and F115W, $\approx$ 1.7 hr for F150W and F356W, and $\approx$ 1.5 hr for F200W and F277W, with 5$\sigma$ AB magnitude limits of $\approx$ 29.0–29.5. The photometry used comes from the AstroDEEP catalog of \citet{Merlin_2024} and the total fluxes were re-estimated in \citet[][]{Castellano_2024} using T-PHOT \citep{TPHOT_2015}. 

\begin{deluxetable}{lcr}\label{tab:params}
\tablecaption{Example BEAGLE-AGN Parameters and Priors}
\tablewidth{\linewidth}
\tablehead{
\colhead{Parameter} &
\colhead{Range} &
\colhead{Prior}
}
\startdata
SFH Parameters& &  \\
\hline
$\tau$ [log$_{10}$(yr)] & [6, 12] & Uniform\\
current\_sfr\_timescale [log$_{10}$(yr)] & [3, 10] & Uniform \\ 
\hline
Galaxy Parameters&  &\\
\hline
redshift & [12.2, 12.4]& Uniform \\
mass [log$_{10}$(M/M$_{\odot}$)] & [5, 12] & Uniform \\
specific\_sfr [yr$^{-1}$] & [$-$14, $-$7] & Uniform \\ 
metallicity [log$_{10}$(Z/Z$_{\odot}$)]  & [$-$2.2, 0.24] & Uniform \\
nebular log$_{10}$(U) & [$-$4, $-$1]  & Uniform\\
nebular log$_{10}$($Z/Z_{\odot}$) & [$-$3, 0.24] & Uniform\\
\hline
Dust Parameters& &  \\
\hline
attenuation type & CF00 &  \\
TauV$_{eff}$& [0.001, 5] & Uniform log$_{10}$\\
nebular $\xi$ & 0.3 & fixed \\
$\mu$ & 0.4  & fixed\\
\hline
AGN Parameters & & \\
\hline
log$_{10}$(L$_{acc}$) &  [40, 47] & Uniform \\
AGN log$_{10}$(U) & [$-$3, 0] &  Uniform \\
AGN log$_{10}$($Z/Z_{\odot}$) &  [$-$3, 0.24] & Uniform \\
AGN $\xi$ & [0.1, 0.5] & Uniform \\
\enddata
\tablecomments{This table outlines the parameters and the corresponding priors used in our \BEAGLE-AGN runs. The dust attenuation model (CF00) are those from \citet{CF_2000}. Nebular $\xi$ is the effective galaxy wide dust-to-metal ratio, AGN $\xi$ is the dust-to-metal ratio in the NLR gas. $\mu$ is the fraction of attenuation optical depth arising from the diffuse ISM. TauV$_{eff}$ is the optical depth of attenuation of the V band. current\_sfr\_timescale is the duration of the current episode of star formation and allows for recent bursts. }
\end{deluxetable}

\section{Methods}\label{sec:methods}

\begin{figure}
    \centering
    \includegraphics[width=\linewidth]{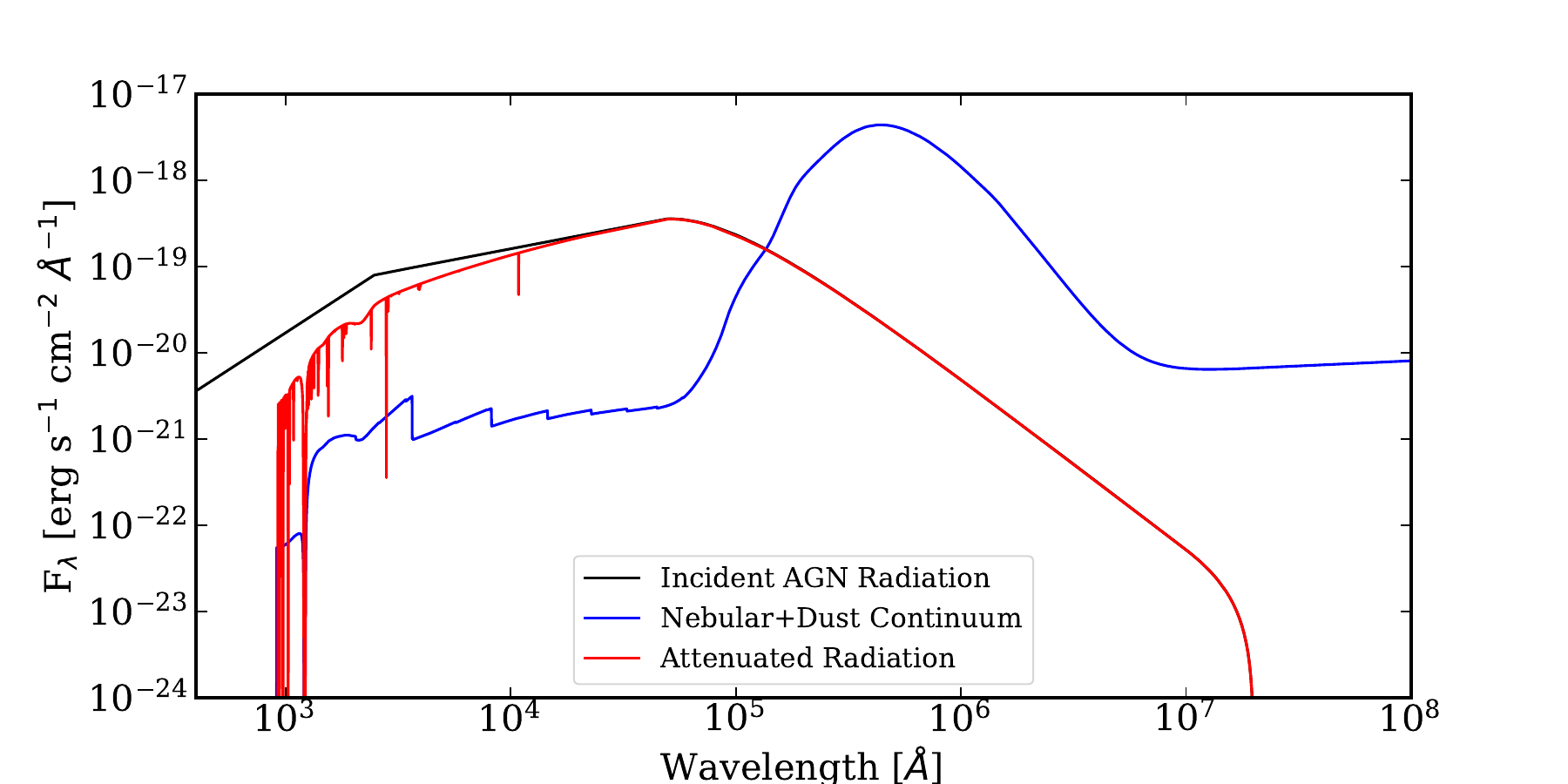}
    \caption{We show the incident AGN spectrum adopted from \citet{Feltre_2016} and its processing through \texttt{Cloudy} as implemented in \BEAGLE-AGN. The black curve represents the incident AGN radiation field, the red curve represents the attenuated incident AGN continuum by the NLR cloud, and the blue curve the nebular+dust continuum emission from the NLR cloud. For clarity, the emission lines have been removed but are included during the fitting procedure to construct the complete model spectrum. Only the blue continuum is used when comparing the model continuum to the observations.}
    \label{fig:incident_agn_spec}
\end{figure}

Decoupling the AGN and stellar components requires simultaneous modeling of both, which we achieve using the \BEAGLE-AGN code \citep{vidal_garcia_2024}, an extension of the \BEAGLE\ framework \citep{chevallard_2016}. \BEAGLE-AGN uses the \texttt{MultiNest} \citep[][]{MultiNest} sampler to explore the multi-dimensional parameter space and determine the best-fitting parameters. 
 
A key advantage of \BEAGLE-AGN is its ability to separate AGN emission in the narrow line region (NLR) from the stellar and \ion{H}{2} contributions. We use this feature to assess the AGN flux contribution to the high-ionization UV lines in GHZ2. \BEAGLE-AGN employs stellar templates based on \citet{BC_2003} stellar population models, with updates described in \citet{vidal_2017}.  For the AGN component, we use updated \citet{Feltre_2016} photoionization grids that incorporate microturbulent velocity effects \citep{Mignoli_microturbulence2019}, as implemented in \citet{vidal_garcia_2024}.
The incident AGN radiation has the following form: 

\begin{equation}
    S_{\nu} \propto \begin{cases}
                    \nu^{\alpha} & 0.001 \leq \lambda/\mu \rm m \leq 0.25 \\
                    \nu^{-0.5} & 0.25 < \lambda / \mu \rm m \leq 10.0 \\
                    \nu^2 & \lambda / \mu \rm m > 10.0
                    \end{cases}
\end{equation}

Within \texttt{Cloudy}, the incident AGN radiation is absorbed, and re-emitted as nebular continuum and emission lines. In the case of a Type-II AGN, we see only the emission from the NLR clouds, since the incident accretion disc emission is hidden behind a dusty torus, as such we only see this post-processed emission through nebular continuum and the emission lines. Figure~\ref{fig:incident_agn_spec} demonstrates the incident spectra (black), the attenuated AGN continuum by the NLR cloud (red) and the post-processed nebular+dust continuum (blue).  We acknowledge that our modeling relies on the strong assumption that the AGN continuum emission is completely obscured by the AGN-heated dust, as an idealized scenario for Type-II AGN. Including this AGN continuum component to the fit would require a full self-consistent radiative transfer modeling of the AGN dust absorption/IR re-remission and scattering. This would also imply additional adjustable parameters. However, this radiative transfer is currently not implemented in \BEAGLE-AGN, a limitation discussed in Section \ref{sec:discussion}.
We also find no evidence that GHZ2 violates this NLR assumption (i.e., no broad-line features in H$\beta$ or H$\alpha$), although the available MIRI/LRS spectroscopy at $R \sim 100$ is not strongly constraining. The AGN power-law slope is fixed to $\alpha = -1.7$, as \citet{vidal_garcia_2024} showed that NLR properties are well recovered for this choice of $\alpha$. To quantify the impact of this assumption, we also generate model grids with $\alpha = -2.0$ and $-1.2$, using the same configuration as the fiducial $\alpha = -1.7$ model. These grids span a range of ionizing spectra from relatively soft to relatively hard and allow us to assess how the assumed AGN slope affects the derived properties. The results of these tests are discussed in Section 4.3. 
  
 \BEAGLE-AGN performs simultaneous spectro-photometric fitting by jointly modeling the observed continuum spectrum, broadband photometry, and, optionally, derived emission-line fluxes. The fitting compares the input spectrum with a physically motivated model spectrum generated from a combination of stellar, nebular, and AGN components, allowing a self-consistent determination of galaxy and AGN properties within a unified Bayesian framework.  This high-resolution model spectrum is then convolved with the NIRSpec line-spread function (LSF), using the instrument resolution curve, to produce the marginal SED that is compared directly to the observed spectrum. Thus, the marginal SED represents the convolved model spectrum used for comparison with the observations. This analysis enables us to characterize the interplay between the AGN and SF components and to explore how different physical assumptions affect the ability of the models to reproduce the spectrum of GHZ2.
 
 To assess the AGN contribution to GHZ2’s nebular emission, we perform simultaneous \BEAGLE-AGN fits to the NIRCam photometry, NIRSpec spectra, and MIRI-derived optical fluxes, using the parameters outlined in Table~\ref{tab:params}. The \BEAGLE-AGN framework does not permit all model parameters to vary simultaneously during the fitting procedure. In particular, the dust-to-metal mass ratio, $\xi$, are fixed within the pre-computed model grids and therefore cannot be treated as free parameters in the Bayesian inference. This limitation is both a consequence of the computational expense associated with constructing sufficiently dense model grids and a deliberate modeling choice adopted during the development of \BEAGLE-AGN.

As demonstrated by \citet{vidal_garcia_2024}, fixing these parameters leads to a more robust recovery of the physically relevant AGN quantities. Their tests on synthetic spectra showed that adopting a fixed value of $\xi=0.3$ yields unbiased estimates of the ionization parameter in the narrow-line region, whereas allowing $\xi$ to vary introduces significant degeneracies. Similarly, $\mu$ was held fixed to avoid propagating uncertainties associated with a parameter that is only weakly constrained by the available emission-line diagnostics. We therefore adopt the same fixed values as \citet{vidal_garcia_2024}, following the validated implementation of the \BEAGLE-AGN framework.

 For stellar-only runs, the AGN parameters were omitted.  Once the fitting is complete, we apply a magnification correction of $\mu = 1.3$ \citep{Zavala_2024} to all derived quantities affected by lensing, such as the stellar mass and star formation rate (SFR).

\begin{deluxetable*}{l l l l}[htbp]
\label{tab:bgl_models}
\tablecaption{Model Descriptions}
\tablehead{
  \colhead{Model} & \colhead{Ionization} & \colhead{Description}
}
\startdata
AGN+SF: Mup100 (fiducial model) & SF+AGN & Default SF+AGN Model as Described in Section~\ref{sec:methods}\\
AGN+SF: Mup300 & SF+AGN & Same as Model 1 but with an SF IMF Cutoff of 300 \Msun \\
SF: Mup100 & SF & Default SF model with an IMF Cutoff of 100 \Msun\\
SF: Mup300 & SF & Same as Model 3 with an IMF Cutoff of 300 \Msun\\
SF: Mup100 + SF C/O & SF & Same as SF: Mup100 with an extended nebular C/O grids ranging from [0.1,  1.2] \\
SF: Mup300 + SF C/O & SF & Same as SF: Mup300 with an extended nebular C/O grids ranging from [0.1,  1.2] \\
SF: Mup100+ $n_H=$10$^4$cm$^{-3}$ & SF & Same as SF: Mup100 with a Hydrogen density of $10^4~\mathrm{cm^{-3}}$ \\
\enddata
\tablecomments{We outline the different \BEAGLE-AGN runs used in this analysis and a short description of the model used. We also have a column that specifies the type of input ionization that went into the model grid generation.}
\end{deluxetable*}

\subsection{\BEAGLE-AGN Models}

 To understand the ionization mechanisms powering GHZ2, we must consider a comprehensive suite of models that incorporate different physical phenomena, as the dominant processes are not known a priori. We employ \BEAGLE-AGN to systematically explore various physical models and identify which combination of stellar populations, nebular physics, and AGN contributions best reproduces the observed spectral features.

 \BEAGLE-AGN produces a family of fiducial models that incorporates both SF and AGN components with the following specifications: the stellar IMF cutoff is 100 \Msun, the nebular log$_{10}$(U) spans $-$4 to $-$1 and the AGN log$_{10}$(U) spans $-$3.5 to $-$0.5, the SF \HII\ regions are modeled at a hydrogen density of 10$^2$ cm$^{-3}$ and the AGN narrow-line regions at 10$^3$ cm$^{-3}$. We added to the fiducial model enhanced N/O SF and AGN grids. These new models are an updated version of the \citet{Feltre_2016} models, which incorporate variations in the N/O abundance ratio. Specifically, the grids are computed for three values of N/O: log$_{10}$(N/O) = $-$1.15 (solar value), $-$0.25, and 0.25 where they are then interpolated in the fitting. These values were chosen to encompass the N/O abundance inferred by \citet{Castellano_2024} and allows flexibility in the N/O abundance. The C/O ratio is fixed to $\log_{10}(\mathrm{C/O})=-0.29$ (i.e., the solar C/O abundance ratio). This value was determined from an initial \BEAGLE-AGN fit in which the C/O ratio was allowed to vary freely and subsequently adopted when generating the enhanced N/O model grid, which requires a fixed C/O abundance. The dust-to-metal mass ratio is fixed to 0.3, and the AGN power-law slope is fixed to $-1.7$. This model is the fiducial SF+AGN model that will be mentioned throughout the paper. Another SF+AGN model using a similar prescription but with the upper mass cutoff of 300 \Msun\ is also explored to test if altering the IMF provides a better fit to the data.

 To explore the SF-only scenario there are several physical scenarios that could plausibly explain the extreme ionization observed in high-redshift galaxies. We therefore carefully consider alternative SF mechanisms that could produce the observed GHZ2 spectral features:

\begin{enumerate}
    \vspace{-2mm}
    \item  Extended stellar IMF: A higher stellar IMF upper-mass cutoff (300 \Msun) would increase the production of hard ionizing photons, potentially boosting high-ionization emission lines. We test models where the SF grids are generated using an IMF cutoff of 300 M$_{\sun}$. These grids have the same parameterization as the base SF-model with the only difference being the upper IMF cutoff. \\
    \vspace{-6mm}
    \item Varying C/O abundances: Super-solar carbon abundances have been measured in other z $>$ 10 sources \citep[][]{Naidu_z14_2025, robertsborsani_civ_ciii_weak_2025, Tang_Highz_properties_2025, Napolitano_2025}, suggesting that enhanced C/O ratios could strengthen \CIV$\lambda$1548+1551 and related features. We test models in which the SF has nebular C/O abundances ranging from 0.1 to 1.2 $(\mathrm{C}/\mathrm{O})/(\mathrm{C}/\mathrm{O}){\odot}$. These grids are generated using the default SF prescription. They are, however, available with different IMF upper-mass cutoffs (100 and 300 $M{\odot}$), both of which we ran. \\
    \vspace{-6mm}
    \item Higher gas densities: Elevated hydrogen densities (n$_H$ $\sim$ 10$^4$ cm$^{-3}$) can boost \CIV $\lambda$1548+1551 via collisional excitation. High-density systems have been observed at high redshift \citep[e.g.,][]{Maiolino_2024}, making this a physically motivated scenario to explore. To test this SF scenario, we run models in which the default SF grids are computed at densities of $10^4$ cm${-3}$ and a fixed $(\mathrm{C}/\mathrm{O})/(\mathrm{C}/\mathrm{O})_{\odot} = 1$. These grids follow the base \BEAGLE-AGN prescription and adopt an IMF upper-mass cutoff of 100 \Msun.
\end{enumerate}

 To ensure completeness, we ran all of the models on the same data so that any differences in the fitting are due to the physical scenarios being explored. A summary of all model types and their relationships are provided in Table ~\ref{tab:bgl_models}.

 To assess model performance, we compute the 50th-percentile marginal SED from \BEAGLE-AGN, weighting each realization by its posterior probability, and then evaluate a weighted $\chi^2$ between this median model spectrum and the NIRSpec spectrum. We choose weights such that $n_{\rm line} , w_{\rm line} = n_{\rm cont} , w_{\rm cont}$, where $n_{\rm line}$ and $n_{\rm cont}$ are the numbers of pixels assigned to emission lines and continuum, and $w_{\rm line}$ and $w_{\rm cont}$ are their respective weights ($w_{\rm line}$/$w_{\rm cont}$ = 6.22). Emission-line pixels are defined as those within $\pm 20$ $\AA$ of the line centers for \ion{N}{4}]$\lambda1486$, \CIV$\lambda1548+1551$, \HeII$\lambda1640$, \ion{O}{3}]$\lambda1660+1666$, \ion{N}{3}]$\lambda1750$, \CIII]$\lambda1908$, \ion{Mg}{2}$\lambda\lambda2795,2802$, \OII$\lambda\lambda3727,3729$, and \NeIII$\lambda3869$.

 Once the weights are determined, we compute the $\chi^2$ of the model spectrum. We also include contributions to $\chi^2$ from the predicted optical and infrared model fluxes to the measured lines, combining all terms into a single overall $\chi^2$. Throughout this work, we quote values from the model with the lowest overall $\chi^2$ and use $\Delta\chi^2$ (relative to this best-fitting model) to assess the relative preference amongst the models.

\begin{figure*}[htp]
    \centering
    \includegraphics[width=.9\linewidth]{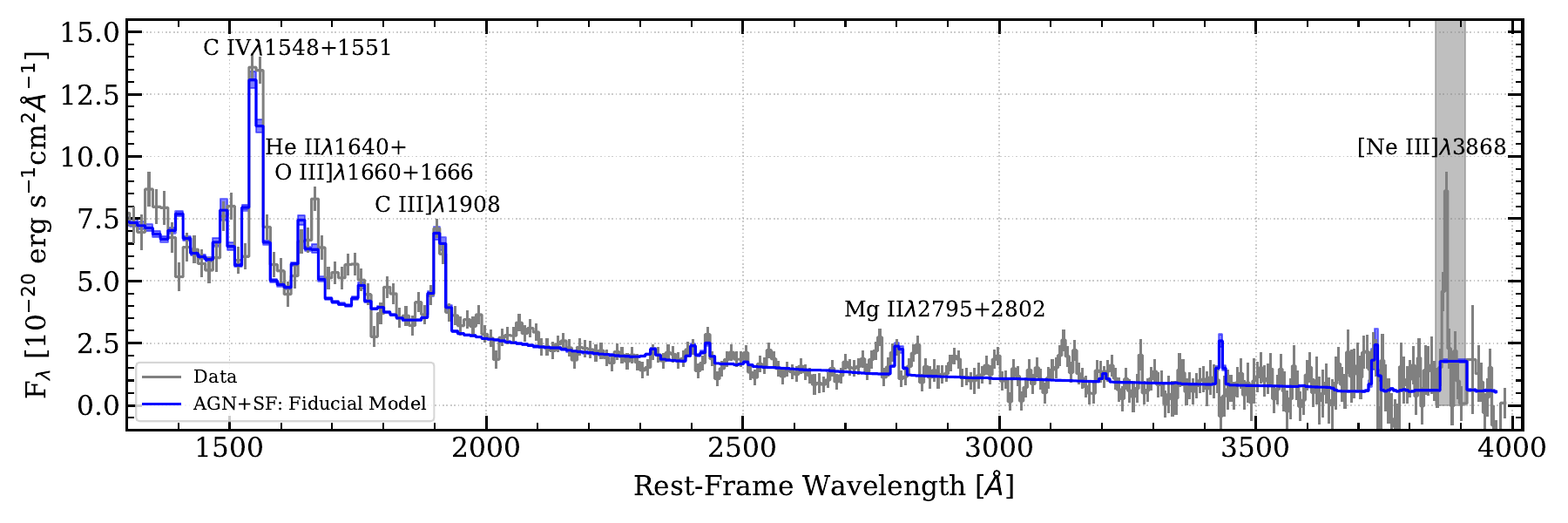}
\end{figure*}
\begin{figure*}[htp]
    \centering
    \includegraphics[width=\linewidth]{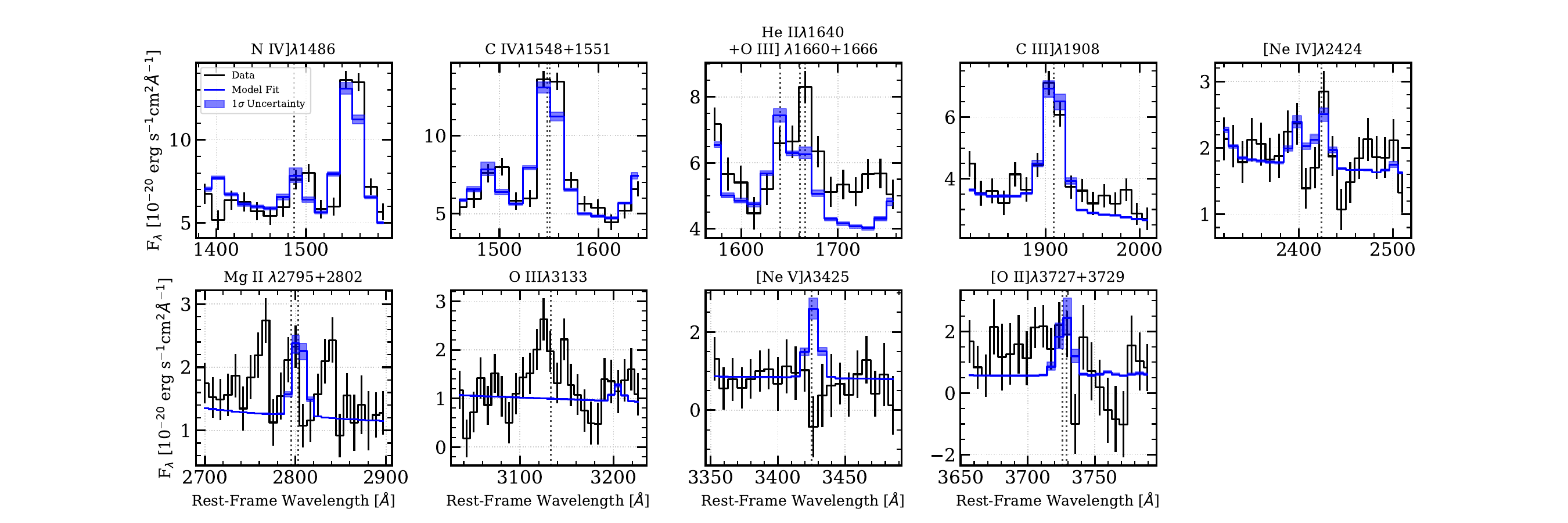}
    \caption{We show the results of our best fit \BEAGLE-AGN fit to the NIRSpec PRISM data and the optical fluxes from the MIRI spectroscopy. The second and third rows show zoom-ins on the emission lines where we compare the model to the data. Note that the \HeII$\lambda$1640 and \ion{O}{3}]$\lambda$1660+1666 are blended in the spectrum and the gray band shows where we masked [\ion{Ne}{3}]$\lambda$3869.}
    \label{fig:fig1}
\end{figure*}

\subsection{Computing the AGN Contribution}\label{sec:agnfraction}

 To quantify the AGN contribution to the emission lines of GHZ2, \BEAGLE-AGN includes built-in extensions that facilitate these calculations. Once the fitting is complete, two extensions in the \BEAGLE-AGN output file, \texttt{AGN EMISSION} and \texttt{HII EMISSION}, are used to compute the fractional AGN contributions to the emission lines. Each row corresponds to a different realization explored by the MultiNest sampler. Each table stores the luminosities of a large suite of emission lines and separates them by emission source (AGN or \ion{H}{2} region). By default, these values are given in units of solar luminosity and are provided in both the emitted and observed frame. For comparison with the GHZ2 observables, we use the observed-frame emission-line columns. The luminosities are first converted to fluxes using the luminosity distance and the fitted redshift, and the \ion{H}{2} and AGN contributions are then combined to obtain the total line flux.

 A direct estimate of the AGN contribution fraction can be obtained by taking the ratio of the AGN emission to the sum of the AGN and \ion{H}{2} emission. However, because MultiNest is not a fully Bayesian sampler, we resample the posterior using the provided weights to ensure proper statistical treatment. \BEAGLE-AGN therefore provides a built-in procedure to compute the AGN contribution fraction for emission lines of interest directly from the posterior samples.

\subsection{Estimating Black Hole Mass}

\BEAGLE-AGN provides a posterior of the accretion luminosity assuming a 10\% covering fraction,  i.e. the fraction of 4$\pi$ steradians covered by the NLR gas, as seen from the accretion disk, which we assume to approximate the bolometric luminosity (log$_{10}$(L$_{acc}$) $\approx$ log$_{10}$(L$_{bol}$)) from the central engine.  We note that this covering fraction is degenerate with the accretion luminosity and imposes a systematic uncertainty on the absolute AGN luminosity tied to the assumed covering fraction, which is not constrained by our data alone.

Under this assumption, we estimate the black hole mass (M$_{BH}$) using the definition of the Eddington ratio as follows:

\begin{equation}\label{eq:M_BH}
    \eta = \frac{L}{L_{Edd}} = \frac{L}{3.2\times 10^4 (\frac{M_{BH}}{M_{\odot}}) L_{\odot}}
\end{equation}

Using the accretion luminosity from \BEAGLE-AGN and Equation~\ref{eq:M_BH}, we estimate black hole masses for a range of Eddington ratios: 0.1, 0.5, and 1, described in \S 4.3.  These masses allow exploration of the co-evolution of the black hole and stellar components via the BH-to-stellar mass relation and to comparison of GHZ2 with other high-redshift AGN in the literature.

 However, this conversion is highly uncertain. For example, our observational constraints on the accretion luminosity come solely from our detected emission lines, which only probe the portion of the spectrum at high enough energies to allow ionization \citep[see~Fig.~1;][]{vidal_garcia_2024}.  The extrapolation to bolometric luminosity is thus heavily dependent on the slope of the accretion disk emission (assumed to be $\alpha$ = $-$1.7 in our work). We approximate the amplitude of this uncertainty by exploring the \citet{Kuboto_agn_models_2019} AGN spectral energy distribution models, finding there to be a scatter of $\sim$0.7 dex in the ratio of the number of ionizing photons to the bolometric luminosity.  We explore and quantify the uncertainty of fixing $\alpha$ = $-$1.7 in Section 4.3. Better constraints on the black hole mass can be obtained via future modeling including the continuum luminosity, and future higher-resolution spectroscopy which could potentially detect broad permitted lines, for an independent kinematic tracer.

\section{Results} \label{sec:results}

 After fitting GHZ2 with a suite of models outlined in Table~\ref{tab:bgl_models}, we present results from our \BEAGLE-AGN analysis. The best fitting model is the model with the lowest overall $\chi^2$ amongst all the models tested (7 total: 2 SF+AGN and 5 SF-only models). Table~\ref{tab:model_comparisons} shows the $\Delta \chi^2$ for all the models and the best fitting model spectrum can be seen in Figure~\ref{fig:fig1}.

\begin{deluxetable}{lc}[htp]
\label{tab:model_comparisons}
\tablehead{
\colhead{Model} & \colhead{$\Delta \chi^2$}
}
\startdata
AGN+SF: Mup100 (fiducial model)&  0\\
AGN+SF: Mup300 &  70.67\\
SF: Mup100 + SF C/O & 1555.06\\
SF: Mup300 + SF C/O &  1605.48\\
SF: Mup300 &  1732.30\\
SF: Mup100 &  1758.55\\
SF: Mup100+ $n_H=$10$^4$cm$^{-3}$ &  1886.78\\
\enddata
\caption{We show the different suite of models used to fit GHZ2 with the difference in $\chi^2$ shown relative to the best fitting model. Models with an AGN component consistently have a lower overall $\chi^2$ values compared with the SF+AGN models. This indicates that an AGN component is necessary to explain the UV spectral features that we see in the NIRSpec spectrum of GHZ2.}
\end{deluxetable}
\vspace{-1cm}

\begin{figure}[htp]
    \centering
    \includegraphics[width=\linewidth]{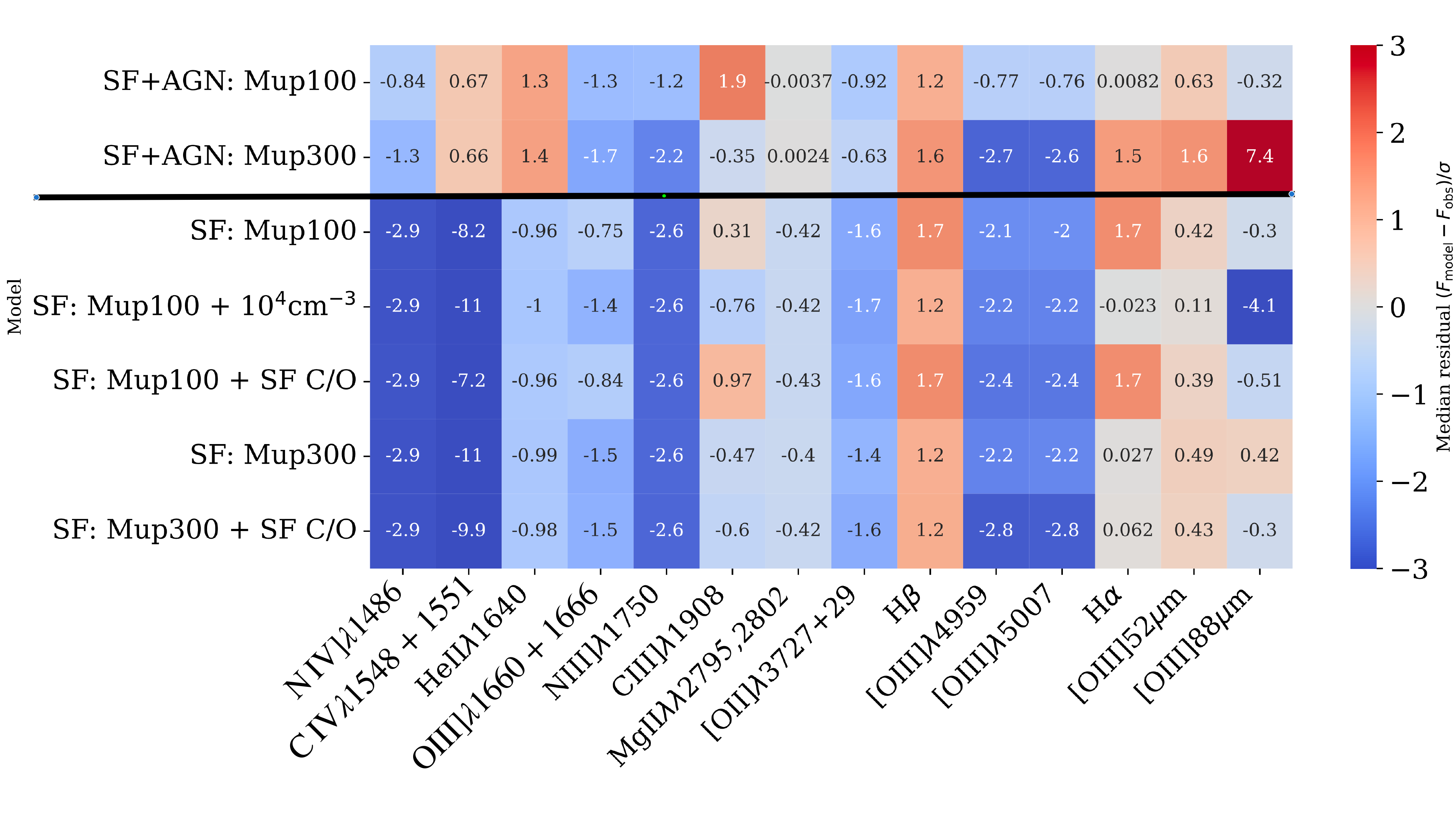}
    \caption{We present median residual heatmaps for line flux predictions across our model suite. Residuals are computed by resampling model fluxes according to their posterior probabilities (Section~\ref{sec:agnfraction}), then subtracting measured fluxes before taking the median. The black line separates SF+AGN from SF-only models. AGN-inclusive models consistently yield lower residuals for high-ionization lines (\CIV$\lambda$1548+1551 and nitrogen lines), indicating that AGN photoionization is required to reproduce the observed emission in GHZ-2. Among all configurations tested, AGN models achieve the lowest residuals across the full line suite. We note that \ion{O}{3}]$\lambda$1660+1666 and \ion{Mg}{2}$\lambda$$\lambda$2795,2802 are blended due to the NIRSpec resolution.}
    \label{fig:flux_comparisons}
\end{figure}

\begin{figure*}
    \centering
    \includegraphics[width=\linewidth]{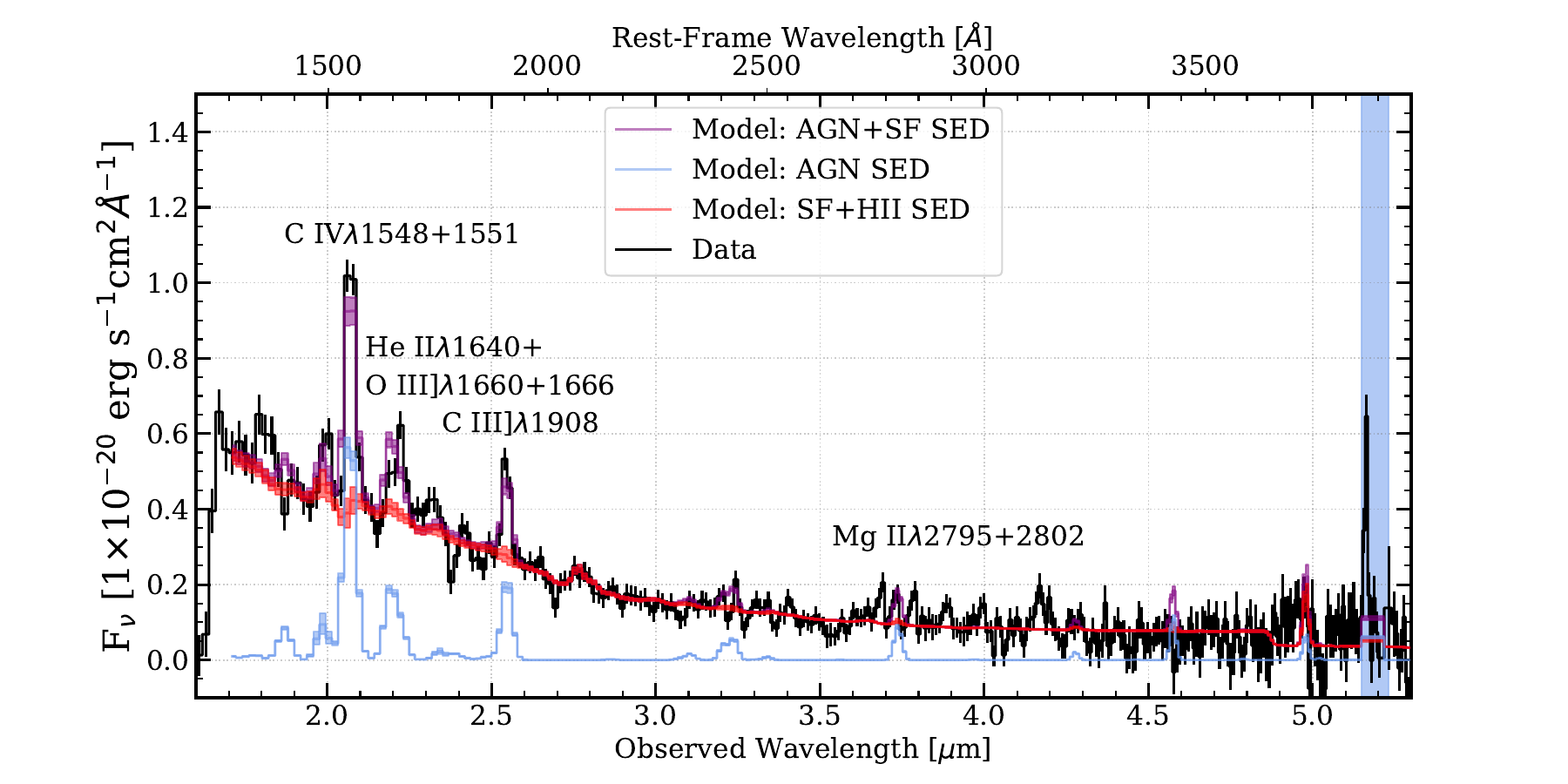}
    \caption{Contribution of the AGN, stellar, and \ion{H}{2} SED components in the best-fit model for GHZ2. The vertical shaded blue region is masked prior to fitting with \BEAGLE-AGN. The continuum shape is dominated by the stellar and \ion{H}{2} components, while the UV emission lines show a non-negligible AGN contribution, most notably for the highest-ionization lines. This modeling therefore supports a partial AGN contribution to the UV emission lines of GHZ2, as evidenced by the strength of these lines in the AGN SED component.}
    \label{fig:Mbh_vs_z}
\end{figure*}

\begin{deluxetable}{ll}[h]
\setlength{\tabcolsep}{20pt} 
\tablecaption{AGN Contribution Fractions with 1$\sigma$ Uncertainties\label{tab:agn_fractions}}
\tablehead{
\colhead{Emission Line} & 
\colhead{AGN Fraction}
}
\startdata
\ion{N}{4}]$\lambda$1486 & 0.89$_{-0.11}^{+0.06}$ \\
\ion{C}{4}$\lambda$1548+1551& 0.97$_{-0.02}^{+0.01}$ \\
\ion{He}{2}$\lambda$1640 & 1.00$_{-0.01}^{+0.00}$ \\
\ion{O}{3}]$\lambda$1660+1666 & 0.78$_{-0.06}^{+0.05}$ \\
\ion{N}{4}]$\lambda$1729 & 1.00$_{-0.01}^{+0.00}$ \\
\ion{N}{3}]$\lambda$1750 & 0.32$_{-0.02}^{+0.01}$ \\
\ion{C}{3}]$\lambda$1908 & 0.64$_{-0.12}^{+0.23}$ \\
\ion{Mg}{2}$\lambda$2795+2802 & 0.92$_{-0.03}^{+0.02}$ \\
$[$\ion{Ne}{3}$]$$\lambda$3869 & 0.61$_{-0.04}^{+0.05}$ \\
H$\beta$ & $<$0.52 \\
$[$\ion{O}{3}$]$$\lambda$4959 & 0.64$_{-0.05}^{+0.05}$ \\
$[$\ion{O}{3}$]$$\lambda$5007 & 0.64$_{-0.05}^{+0.06}$ \\
H$\alpha$ & 0.45$_{-0.05}^{+0.06}$ \\
$[$\ion{O}{3}$]$52$\mu$m & 0.52$_{-0.07}^{+0.08}$ \\
$[$\ion{O}{3}$]$88$\mu$m & 0.29$_{-0.05}^{+0.07}$ \\
\enddata
\tablecomments{AGN contribution fractions estimated for different emission lines with asymmetric 1$\sigma$ uncertainties. We note that the due to the spectral resolution of the NIRSpec prism the \ion{He}{2}$\lambda$1640 and \ion{O}{3}]$\lambda$1660+1666 and the \ion{Mg}{2}$\lambda$$\lambda$2795,2802 are blended. We provide the 84th percentile upper limit of the AGN fraction for H$\beta$.}
\end{deluxetable}

\subsection{Best Fitting Model}

 We find that the best fitting model is the fiducial SF+AGN model with the enhanced N/O grids. Figure~\ref{fig:fig1} shows the best fitting model with respect to the NIRSpec spectrum on top, and zoom-in panels on specific emission lines on the bottom, showing excellent agreement with the observed data. While the model is able to reproduce most lines, it under-predicts the strength of \ion{N}{3}]$\lambda$1750 but is within the 1$\sigma$ error of the measured values, and the weaker \ion{O}{3}$\lambda$3133 model flux is 4$\sigma$ under predicted as seen in Figure~\ref{fig:fig1}. The model also over-predicts the [\ion{Ne}{5}]$\lambda$3425 line by over 7$\sigma$ relative to the noise, as the spectrum has no measurable [\ion{Ne}{5}]$\lambda$3425 emission. Another line that is slightly underpredicted by the best-fitting model is \ion{O}{3}]$\lambda$1660+1666, with the predicted flux falling approximately 1$\sigma$ below the measured value. Given the observational uncertainties, this difference is not statistically significant, and the model provides an overall excellent match to the observed UV emission-line spectrum. 
 
One of the lines that is seen in the cutouts is the \ion{S}{4}$\lambda$1394,1403 doublet in the far ultraviolet in the upper left panel. This line is tentatively seen in absorption and its origin could come from hot stars, however the best fitting models has this feature seen in emission rather than absorption. However, the significance of the line is consistent with noise fluctuations. Another limitation for this absorption line is that the NIRSpec prism resolution is not high enough to properly study the absorption in detail. Another line for which we find a discrepancy between the preferred model and the observations is the [\ion{Ne}{5}]$\lambda3426$ emission line. The fiducial SF+AGN model predicts detectable [\ion{Ne}{5}] emission, whereas no significant emission is observed. We note that [\ion{Ne}{5}]$\lambda3426$ was not included as an observational constraint in the fitting procedure, and therefore its predicted strength represents an independent prediction of the preferred photoionization model rather than a quantity explicitly optimized during the fit. In addition, [\ion{Ne}{5}]$\lambda3426$ is among the highest-ionization optical emission lines, requiring photons with energies exceeding 97 eV. Its strength is therefore particularly sensitive to the extreme-UV portion of the AGN ionizing spectrum. In the adopted \BEAGLE-AGN model grids, the AGN ionizing continuum is fixed to the spectral prescription of \citet{Feltre_2016}, including a power-law slope of $\alpha=-1.7$, and is not treated as a free parameter. Consequently, the overprediction of [\ion{Ne}{5}] likely reflects limitations of the adopted photoionization model or its assumptions regarding the ionizing continuum, rather than the fitting procedure itself. While this discrepancy represents an important caveat, the SF+AGN model nevertheless provides the best overall agreement with the ensemble of continuum and emission-line constraints included in the likelihood.
 
Another feature that we see around the Balmer jump is a slight drop in the continuum level. Although the BEAGLE-AGN model predicts a Balmer jump, this feature should be interpreted as a model-dependent consequence of the full SED fit rather than as a direct spectroscopic detection. The continuum near rest-frame 3646$\AA$ is weak, so the constraint on the break is driven primarily by the broadband photometry and by the assumed stellar and nebular continuum components. Physically, the inferred jump may reflect the contribution of nebular bound-free continuum emission, or the need to reproduce the observed photometric colors redward of the break. We therefore treat the Balmer jump as suggestive evidence for the continuum shape implied by the best-fitting model, while noting that its strength remains sensitive to the adopted star-formation history, and nebular continuum treatment.

 We show the best fitting model parameters and their derived flux values in Table~\ref{tab:combined_results}. The stellar mass and SFR have been corrected by the magnification factor of 1.3 as measured in \citet{Zavala_2024}. Some of the properties we infer are: stellar-mass log$_{10}$(M$_{*}$/M$_{\odot}$) = $8.41^{+0.21}_{-0.13}$, and accretion luminosity log$_{10}$(L$_{acc}$/erg s$^{-1}$) = $44.98^{+0.03}_{-0.03}$ which is used to infer a black hole mass. We also compute the thermal re-radiation from dust absorption of the best fitting model by integrating the best fitting model from 8 - 1000 $\mu$m and we get that the thermal re-radiation from dust absorption flux is: 3.81$_{-0.62}^{+0.62} \times 10^{-18}$ erg s$^{-1}$cm$^{-2}$ and can be used to compare against future IR measurements.

 To test whether the line fluxes and spectrum of GHZ2 are better reproduced by an alternative SF+AGN model, we examine the effect of increasing the upper-mass cutoff of the stellar IMF to 300 $M_{\odot}$. We find that this additional SF+AGN model does not improve the fit, with a $\Delta\chi^2 = 70.67$, statistically disfavoring the stellar IMF upper-mass cutoff of 300 $M_{\odot}$. This indicates that an extremely top-heavy IMF with a 300 $M_{\odot}$ cutoff is neither statistically nor physically favored for GHZ2.

 Comparing the SF+AGN models to the SF-only models, we find that the latter fail to capture the strengths of the high-ionization emission lines seen in GHZ2. The main drivers are \CIV$\lambda1548+1551$, \ion{N}{4}]$\lambda1486$, and \ion{N}{3}]$\lambda1750$, which all SF-only models fail to reproduce, as shown in Figures~\ref{fig:flux_comparisons} and \ref{fig:bp_comparison}. For example, comparing the best-fitting SF+AGN model with the best-fitting SF-only model yields a $\Delta\chi^2 = 1555.06$, which strongly disfavors the SF-only scenario. Even allowing for variable C/O abundance, in an attempt to boost \CIV$\lambda1548+1551$, is not sufficient to favor the SF-only hypothesis. Our modeling strongly supports the presence of AGN ionization to explain the high-ionization emission lines.

 We explored whether elevated nebular densities ($n_H = 10^4$ cm${-3}$) could enhance \CIV$\lambda1548+1551$ emission through collisional excitation in SF-only models. However, this configuration performs worse than the baseline SF-only model and still fails to match the observed \CIV$\lambda1548+1551$ and nitrogen line strengths. This is further quantified by the $\Delta \chi^2 = 1886.78$, making this a statistically disfavored model from the SF+AGN scenario. All tests indicate that density effects alone cannot account for GHZ2's high-ionization emission and that other mechanisms are required.

 We also explore the SF-only model with an extended IMF cutoff of 300 \Msun\ and that fails to reproduce GHZ-2's observables with a $\Delta\chi^2 = 1732.30$ relative to the best fitting SF+AGN model. Despite the increased production of ionizing photons from massive stars, the model significantly underestimates high-ionization lines like \CIV$\lambda$1548+1551 and \HeII$\lambda$1640 (Figures~\ref{fig:flux_comparisons} and \ref{fig:bp_comparison}). This suggests that stellar photoionization alone—even with a top-heavy IMF—is insufficient, and a harder ionizing spectrum from an AGN is necessary to explain the observations.

As a last test of the SF-only scenario, we follow the approach of \citet{robertsborsani_civ_ciii_weak_2025}, who showed that high-redshift galaxies with strong \CIV$\lambda$1548+1551 are likely observed during a recent star formation burst within the last 3 Myr. We perform an independent run of the (stellar-only) \texttt{Bagpipes} SED-modeling code with a non-parametric star formation history to assess whether a recent burst could reproduce the \CIV$\lambda$1548+1551 emission. We adopt the BPASS 2.2.1 models \citep{bpass_221_2018} with an extended log$_{10}$(U) grid (–4 to 0.5) and two cloud densities (10$^2$ and 10$^4$ cm$^{-3}$), using the non-parametric SFH framework of \citet{Leja_2019} with priors allowing for burstiness. The SFH bins are [0, 3, 10, 25, 50] Myr, extended with five additional logarithmically spaced bins up to the age of the universe at GHZ2’s redshift of $z$ = 12.34. We implement a flexible dust attenuation model following \citet{Salim_2018}, fitting both photometry and NIRSpec spectra, while removing filters blueward of the Lyman break and using the masked spectra from the \BEAGLE-AGN fit for consistency between SED codes.

This \texttt{Bagpipes} run does favor a recent star formation burst within the last 3 Myr; however, this burst is insufficient to reproduce the observed \CIV$\lambda$1548+1551 strength in GHZ2. The high-density (10$^4$ cm$^{-3}$) \texttt{Bagpipes} run reproduces only the continuum level flux at rest-frame $\lambda$ = 1550 and there is no flux contribution of the SF-model to \CIV$\lambda$1548+1551. Figure~\ref{fig:bp_comparison} compares the \texttt{Bagpipes}, \BEAGLE\ SF-only runs and the SF+AGN \BEAGLE\ runs, and highlights that SF-only cannot reproduce \CIV$\lambda$1548+1551, while \CIII]$\lambda$1908 is successfully modeled by all the SF models. This suggests that a harder ionizing spectrum is required to populate carbon to the \CIV\ state, implying that an AGN is a likely source of the necessary ionizing radiation.

\begin{deluxetable*}{lrr|lrr|lrr}[htp]
\tablecaption{Best-fit Parameters and Predicted Observed Fluxes from \BEAGLE-AGN\label{tab:combined_results}}
\tablewidth{0pt}
\tablehead{
\multicolumn{3}{c|}{Best-fit Parameters} & \multicolumn{3}{c|}{UV-Optical Fluxes} & \multicolumn{3}{c}{IR Fluxes} \\
\hline
\colhead{Parameter} & \colhead{Median} & \colhead{Max. Like.} & 
\colhead{Line} & \colhead{Flux$^a$} & \colhead{Res} & 
\colhead{Line} & \colhead{Flux$^b$} & \colhead{Res}
}
\startdata
$\tau$ & $9.98^{+1.39}_{-1.41}$ & $9.69$ & \ion{N}{4}] $\lambda1486$ & $0.55_{-0.12}^{+0.13}$ & -0.83 & $[$\ion{O}{3}$]$ 51$\mu$m & $0.35_{-0.03}^{+0.04}$ & 0.63 \\
$\log_{10}(Z/Z_\odot)$ & $-1.91^{+0.41}_{-0.20}$ & $-1.96$ & \ion{C}{4} $\lambda\lambda1548,1551$ & $2.71_{-0.12}^{+0.14}$ & 0.71 & $[$\ion{O}{3}$]$ 88$\mu$m & $0.28_{-0.04}^{+0.04}$ & -0.31\\
$\log_{10}(M_*/M_\odot)$ & $8.41^{+0.21}_{-0.13}$ & $8.32$ & \ion{He}{2} $\lambda1640$ & $0.76_{-0.04}^{+0.04}$ & 1.31 & -- & -- & --\\ 
$\log_{10}(\mathrm{Age}/\mathrm{yr})$ & $7.72^{+0.33}_{-0.20}$ & $7.59$ & \ion{O}{3}] $\lambda\lambda1660,1666$ & $0.39_{-0.02}^{+0.03}$ & -1.30 & -- & -- & --\\ 
$\log_{10}(\mathrm{sSFR}/\mathrm{yr}^{-1})$ & $-10.39^{+2.60}_{-2.54}$ & $-8.57$ & \ion{N}{3}] $\lambda1750$ & $0.30_{-0.10}^{+0.10}$ & -1.20 & -- & -- & --\\ 
$\log_{10}(t_{\mathrm{SFR}}/\mathrm{yr})$ & $4.50^{+1.13}_{-0.98}$ & $4.37$ & \ion{C}{3}] $\lambda\lambda1908$ & $1.11_{-0.06}^{+0.07}$ & 1.90 & -- & -- & --\\ 
$\log_{10}(U_{\mathrm{neb}})$ & $-2.04^{+0.15}_{-0.14}$ & $-2.03$ & \ion{Ne}{4}] $\lambda2424$ & $0.15_{-0.02}^{+0.02}$ & -- & -- & -- & --\\ 
$\log_{10}(Z_{\mathrm{neb}}/Z_\odot)$ & $-0.84^{+0.15}_{-0.14}$ & $-0.82$ & \ion{Mg}{2} $\lambda\lambda2795,2802$ & $0.26_{-0.02}^{+0.02}$ & -0.0036 & -- & -- & --\\
$\tau_{V,\mathrm{eff}}$ & $0.15^{+0.03}_{-0.03}$ & $0.15$ & [\ion{Ne}{5}] $\lambda3426$ & $0.21_{-0.03}^{+0.03}$ & --  & -- & -- & --\\ 
$\log_{10}(L_{\mathrm{acc}}/\mathrm{erg\,s^{-1}})$ & $44.98^{+0.03}_{-0.03}$ & $44.98$ & [\ion{O}{2}] $\lambda\lambda3727,3729$ & $0.29_{-0.05}^{+0.06}$ & -0.93 & -- & -- & --\\ 
$\log_{10}(U_{\mathrm{AGN}})$ & $-1.97^{+0.09}_{-0.06}$ & $-2.00$ & [\ion{Ne}{3}] $\lambda3869$ & $0.54_{-0.03}^{+0.03}$ & -1.50 & -- & -- & --\\ 
$\log_{10}(Z_{\mathrm{AGN}}/Z_\odot)$ & $-0.65^{+0.09}_{-0.09}$ & $-0.67$ & H$\beta$ $\lambda4861$ & $< 0.97$ & 1.20  & -- & -- & --\\ 
-- & -- & -- & $[$\ion{O}{3}$]$ $\lambda4959$ & $1.64_{-0.09}^{+0.09}$ &-0.77 & -- & -- & --\\ 
-- & -- & -- & $[$\ion{O}{3}$]$ $\lambda5007$ & $4.94_{-0.27}^{+0.28}$ & -0.75  & -- &  -- & --\\
-- & -- & -- & H$\alpha$ $\lambda6563$ & $2.60_{-0.25}^{+0.20}$  & -0.01 & -- &  -- & --\\
\enddata
\tablenotetext{a}{Units: $10^{-18}$ erg s$^{-1}$ cm$^{-2}$}
\tablenotetext{b}{Units: $10^{-20}$ erg s$^{-1}$ cm$^{-2}$}
\tablecomments{Best-fit parameters (median and 16th–84th percentile uncertainties, plus maximum likelihood values) and predicted emission line fluxes from the best-fitting \BEAGLE-AGN model. The modeled uncertainty from the BEAGLE-AGN  represent the uncertainty on the inferred observed line fluxes within the adopted model framework, rather than the measurement uncertainties associated with the individual spectral features, see Table~\ref{tab:obs_flux} for the observed fluxes. Res is the residual which is measured using (f$_i$($\theta$) - f$_i$)/$\sigma_i$, where i corresponds to a specific emission line and $f_i(\theta)$ is the predicted model flux. }
\end{deluxetable*}

\subsection{AGN Contribution Fraction}

After fitting the models, we employed the methods outlined in Section~\ref{sec:agnfraction} to compute the AGN contribution fraction for key emission lines. The results are compiled in Table~\ref{tab:agn_fractions} and indicate AGN fractional contributions of $97_{-2}^{+1}\%$ for \CIV$\lambda1548+1551$ and $64_{-12}^{+23}\%$ for \CIII]$\lambda1908$. Table~\ref{tab:agn_fractions} also shows the AGN contribution to other UV and optical emission lines; as expected, the highest-ionization lines exhibit the largest AGN contribution fractions. Most notably, He II$\lambda1640$ and the nitrogen line \ion{N}{4}$\lambda1486$ both show AGN contributions $\geq 85\%$. We also find non-negligible AGN contributions to other emission lines, such as \OIII$\lambda5007$, with values of $64_{-5}^{+6}\%$, respectively, implying that a substantial AGN contribution powers the UV and optical emission lines of GHZ2.

\subsection{Black Hole Mass}

Using the accretion luminosity ($L_{acc}$) posterior from the best-fitting \BEAGLE-AGN run, we infer the black hole mass (M$_{BH}$) via Equation~\ref{eq:M_BH}, 
assuming Eddington ratios of 1, 0.5, and 0.1. The median black hole masses computed are $\log_{10}$(M$_{BH}$/\Msun) = 6.89$^{+0.03}_{-0.03}$, 7.19$^{+0.03}_{-0.03}$, 7.89$^{+0.03}_{-0.03}$.  To test the impact of the AGN power-law slope on the inferred black hole mass, we generated model grids using the same configuration as the fiducial model, but with $\alpha$ fixed to $-2.0$ and $-1.2$ (compared to the fiducial value of $-1.7$). These alternative $\alpha$ grids provide poorer fits to the data, as quantified by their $\Delta\chi^2$ values relative to the fiducial model: $\Delta\chi^2 = 6$ for $\alpha = -2.0$ and $\Delta\chi^2 = 43$ for $\alpha = -1.2$.

 To assess the systematic uncertainty associated with our assumption about the AGN power-law slope, $\alpha$, we repeated the full modeling for each choice of $\alpha$ and compared the resulting posterior distributions. For each $\alpha$ model, we took the median of the posterior for the black hole mass, and we quantified the systematic uncertainty as the difference between these median values to the median of the fiducial $\alpha$ = $-$1.7 model. In the case of the black hole mass, the average shift in the posterior medians between the different $\alpha$ assumptions relative to the $\alpha$ = $-$ 1.7 corresponds to a systematic uncertainty of $0.6$ dex. Regardless of this systematic uncertainty, the inferred black hole masses remain comparable to those of other high-redshift AGN.

\begin{figure}[htp]
    \centering
    \includegraphics[width=\linewidth]{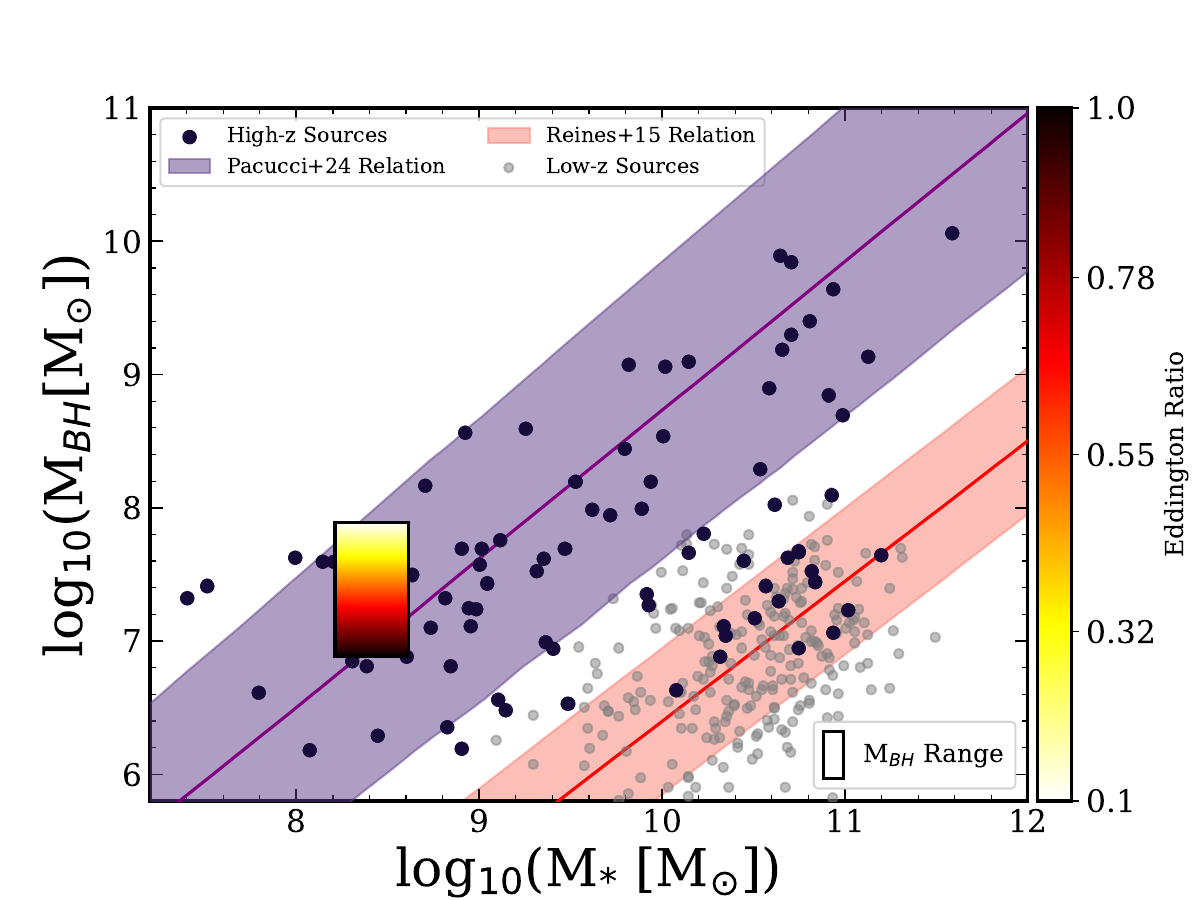}
    \caption{We plot the black hole-to-stellar mass of high redshift AGN from the literature \citep[to cite a few:][]{Greene_2024, Harikane_agn_survey_2023, Kocevski_2024, Matthee_2024, Taylor_A_2024, hollis_bh_mass_2025}, compared to our estimates for GHZ2. The black rectangular gradient shows the range of black hole masses from 0.01 to 1, consistent with the high-z relation. For the ratio of BH-to-stellar mass we measure a M$_{BH}$/ M$_{*}$ = 0.03$^{+0.01}_{-0.01}$, 0.06$^{+0.02}_{-0.02}$, and  0.30$^{+0.09}_{-0.11}$ assuming an $\eta$ = 1, 0.5, 0.1 for GHZ2. This result shows that GHZ2 harbors an overmassive black hole compared to the local relation \citep{Reines_2015}, consistent with other high redshift {\it JWST}-discovered AGN and the high-z relation measured by \citet{Pacucci_2024}.}
    \label{fig:mbh_mstar_ratio}
\end{figure}

\subsection{Black Hole-Stellar Mass Relation}

\BEAGLE-AGN enables the decoupling of AGN and stellar components, allowing analysis of GHZ2’s black hole-to-stellar mass ratio. 
Using the stellar and black hole mass from the best fitting model, we find an elevated M$_{BH}$/M$_*$ compared to the local relation, consistent with high-redshift AGN, measuring M$_{BH}$/M$_*$ = 0.03$^{+0.01}_{-0.01}$, 
0.06$^{+0.02}_{-0.02}$, and  0.30$^{+0.09}_{-0.11}$, for $\eta$ = 1, 0.5, 0.1 respectively (Figure~\ref{fig:mbh_mstar_ratio}; \citealt{Pacucci_2023,Harikane_2023,Ubler_2023, Ding_2023,Pacucci_2024, Direct_BH_Mass_2025}). If the AGN interpretation holds, GHZ2 provides a stringent test case for black hole seeding and co-evolution models to explore how a system can reach the M$_{BH}$/M$_*$ ratio by $z$ = 12.34.

\section{Discussion} \label{sec:discussion}

\citet{Castellano_2024} and \citet{Calabro_2024} conducted a detailed analysis of GHZ2 to investigate the origins of its UV emission. They employed line ratio diagnostics involving \CIV$\lambda$1548+1551, \CIII]$\lambda$1908, and \HeII$\lambda$1640, comparing the observed ratios to predictions from AGN models \citep{Feltre_2016} and stellar models \citep{Gutkin_2016}. A key limitation of these model grids is the assumption that the ionizing spectrum is dominated by a single source—either AGN or stellar. Accurately characterizing the UV ionization in GHZ2, therefore, requires a framework capable of incorporating a multi-component ionizing source. The \BEAGLE-AGN approach addresses this limitation, as demonstrated by \citet{vidal_garcia_2024}, by simultaneously modeling stars, \HII\ regions, and the narrow-line region (NLR) to reproduce nebular emission lines.  Additionally, our approach models individual line-strengths, rather than just the line ratio.

\subsection{Model Comparisons}

After exploring a wide range of physical scenarios for both SF+AGN and SF-only models, we find that the fiducial SF+AGN model with an IMF upper-mass cutoff of 100  \Msun is the most statistically favored.  This model simultaneously reproduces the UV continuum shape and matches a suite of UV and optical emission lines, indicating that it has (i) the appropriate mix of stellar populations to power the observed continuum and (ii) the correct balance of AGN and stellar ionizing contributions to reproduce the nebular line spectrum.

 Variations in the IMF upper-mass cutoff (100 vs. 300  \Msun ) do measurably affect the fits. In particular, the 300  \Msun\ SF+AGN model is strongly disfavored, with a $\Delta \chi^2$=75.19 relative to the 100  \Msun\ model. This discrepancy is driven primarily by the optical emission-line fluxes: models with an upper-mass cutoff of 300  \Msun\ systematically fail to capture the strengths of the optical nebular lines, whereas the 100  \Msun\ models reproduce these fluxes within the observational uncertainties.

 The lack of model agreement to the optical line emission in the 300 \Msun\ models is a direct consequence of their harder ionizing spectra. Extending the IMF to higher masses increases the contribution from very massive stars ($\gtrsim 100,M_{\odot}$), which substantially boosts the flux of high-energy ionizing photons (e.g., He-ionizing photons) while only modestly changing the total H-ionizing photon budget \citep[e.g.,][]{stanway_mup_ionization_2019}. The fact that the 300 \Msun\ models cannot predict these optical line fluxes, while the 100  \Msun\ models do not, therefore provides direct evidence that an IMF extending to 300 \Msun\ would produce an ionizing spectrum that is too hard for this system.

 Taken together, the ability of the 100  \Msun\  SF+AGN model to reproduce both the UV continuum and the UV+optical emission-line constraints, combined with the statistically significant degradation in fit quality for the otherwise identical 300  \Msun\  model, leads us to conclude that an IMF upper-mass cutoff of 100  \Msun\ is strongly preferred over the 300\Msun\ for this source. 

 To understand why the SF-only models fail to reproduce GHZ2, we examine the physics of the baseline SF-only configuration with an IMF upper-mass cutoff of 100 \Msun\ This model successfully matches the observed UV continuum shape but fails to reproduce the high-ionization emission lines. This indicates that while stellar populations dominate the continuum emission in GHZ2, an additional, harder ionizing component is required to power the observed nebular line spectrum—consistent with an AGN-like contribution.

 The inability of the baseline SF-only model to generate the necessary high-ionization line fluxes reflects the intrinsic limitations of its ionizing spectrum: even with a realistic stellar IMF, the stellar radiation field is not sufficiently hard to account for the observed excitation state of the gas. Motivated by this mismatch, we systematically explore alternative SF-only scenarios (e.g., varying IMF, metallicity, and gas conditions) to test whether any purely stellar configuration can simultaneously reproduce both the continuum and emission-line properties of GHZ2.

 Another point of comparison within the SF-only models is the inclusion of a variable C/O abundance. Even when we allow C/O to vary, the SF-only grids still fail to reproduce the spectrum of GHZ2. These models systematically predict a lack of strong \CIV$\lambda1548+1551$ emission observed in GHZ2, while also underpredicting the \HeII$\lambda1640$ and \ion{O}{3}]$\lambda1660+1666$ features.

 From a physical standpoint, increasing C/O primarily raises the gas-phase carbon abundance and can enhance carbon lines such as \CIII]$\lambda1908$ and \CIV$\lambda1548+1551$ relative to oxygen lines at fixed O/H. However, strong nebular \CIV$\lambda1548+1551$ emission additionally requires a very hard ionizing spectrum capable of maintaining a substantial fraction of C$^{3+}$, as well as a high ionization parameter \citep[][]{izotov_civ_2024}. In our SF-only models, even with elevated C/O, the stellar radiation field remains too soft to generate the required C$^{3+}$ and He$^{+}$-ionizing photon fluxes. As a result, \CIV$\lambda1548+1551$ and \HeII$\lambda1640$ are both under produced, and the high-excitation \ion{O}{3}]$\lambda1660+1666$ lines are likewise too weak.

 Thus, allowing for enhanced C/O abundance does not resolve the discrepancies: carbon enrichment can amplify carbon lines once a sufficiently hard ionizing spectrum is present, but it cannot compensate for the lack of hard photons in the SF-only models. Stellar emission alone can reproduce some aspects of the UV continuum and the \CIII]$\lambda1908$ emission, but it cannot provide the ionization conditions necessary to account for the strong \CIV$\lambda1548+1551$ and associated high-ionization features observed in GHZ2.

We also test SF-only models with higher hydrogen densities and find that these are likewise unable to match the high-ionization lines seen in GHZ2. Moreover, within the density range probed by our SF-only models (10$^4$cm$^{-3}$), UV semi-forbidden lines such as \CIII]$\lambda1908$ and \ion{O}{3}]$\lambda\lambda1660,1666$ have critical densities of order 10$^9$ -- 10$^{10}$ cm$^{-3}$ and therefore remain in the low-density regime, where collisional de-excitation is negligible \citep[e.g.,][]{Osterbrock_2006, zheng_ncrit_1988, Nakajima_2018}. As a result, varying the gas density primarily affects low–critical-density lines (e.g. \OII, \SII) rather than boosting the high-ionization UV features that are under predicted in the SF-only models.

Thus, while density variations can modestly influence emission-line ratios through changes in log(U) and collisional quenching of low–critical-density lines, they are not sufficient—within SF-only models—to generate the high-ionization emission observed in GHZ2 due to a lack of hard ionizing spectrum.

GHZ2 provides a valuable test case because line diagnostics place it in a region of parameter space intermediate between AGN and star formation, suggesting that a combination of both is required to explain its spectral features. The \BEAGLE-AGN SED fitting supports this scenario. \BEAGLE-AGN's self-consistent SED modeling enables us to assess evidence for AGN activity, as line diagnostics alone are limited by degeneracies in the ionization mechanisms \citep{Calabro_2024, Castellano_2024, Cleri_2025} and we further quantify the extent of the AGN contribution.

\begin{figure}
    \centering
    \includegraphics[width=\linewidth]{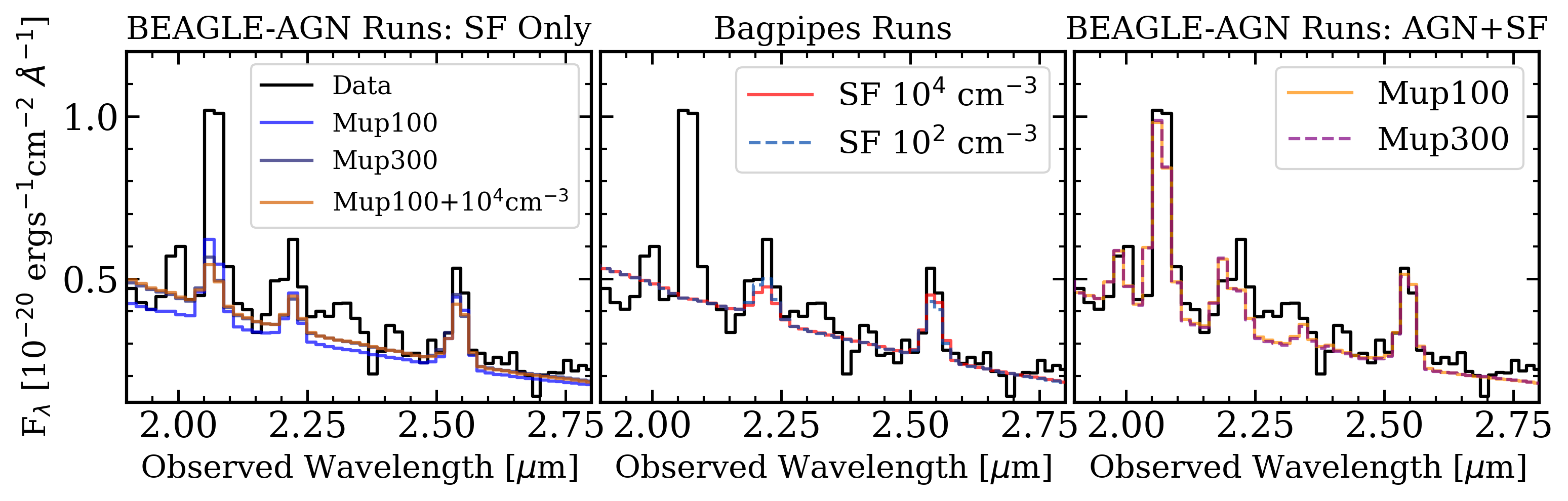}
    \caption{A model comparison of the SF BEAGLE models, the SF Bagpipes models and the best fitting AGN+SF models from BEAGLE. We see that the SF only models are not able to match either: i) the UV continuum slope as evident in the \BEAGLE-AGN models or ii) the strength of the \CIV$\lambda$1548+1551 emission or both. The only models that are able to match the data are models with an AGN component, as they simultaneously reproduce both the UV slope and the strength of \CIV$\lambda$1548+1551.}
    \label{fig:bp_comparison}
\end{figure}

\subsection{Testing Black Hole Seeding}

If the AGN interpretation is correct, GHZ2 would host the earliest supermassive black hole (SMBH) observed to date. Such early SMBH formation would provide an opportunity to push theoretical models of black hole seeding and growth to their limits \citep{Smith_Bromm_2019, Woods_2019, Inayoshi_2020}. In general, to promote the accelerated emergence of SMBHs, one can consider heavy initial seed masses, earlier redshifts of formation, or more efficient growth through super-Eddington accretion. It is also conceivable that a combination of all these effects may be required to understand the black hole growth in GHZ2. 

One leading scenario invokes high-mass seed black holes, with initial masses of $\sim 10^{4-6}$\,M$_{\odot}$, as opposed to `light seed' stellar remnants with $\lesssim$~100~M$_{\odot}$. Such `heavy' seeds could form via the direct collapse of a massive pristine gas cloud, where vigorous fragmentation into a Population~III star cluster is suppressed by keeping the gas temperature close to the (hydrogen) atomic cooling threshold \citep[e.g.,][]{Bromm_2003, Begelman_2006,Lodato_2006}. These direct-collapse black hole (DCBH) models, however, require conditions that may be quite rare, such as the presence of strong soft (Lyman-Werner) UV fluxes \citep[e.g.,][]{Habouzit_2016, Bhowmick_2024}. Other pathways which can form heavy seeds may also be in play, involving dynamical processes in dense star clusters, such as the merging of stars or their BH remnants \citep[e.g.,][]{Reinoso_2023,Kritos_2024}.

\begin{figure}
    \centering
    \includegraphics[width=\linewidth]{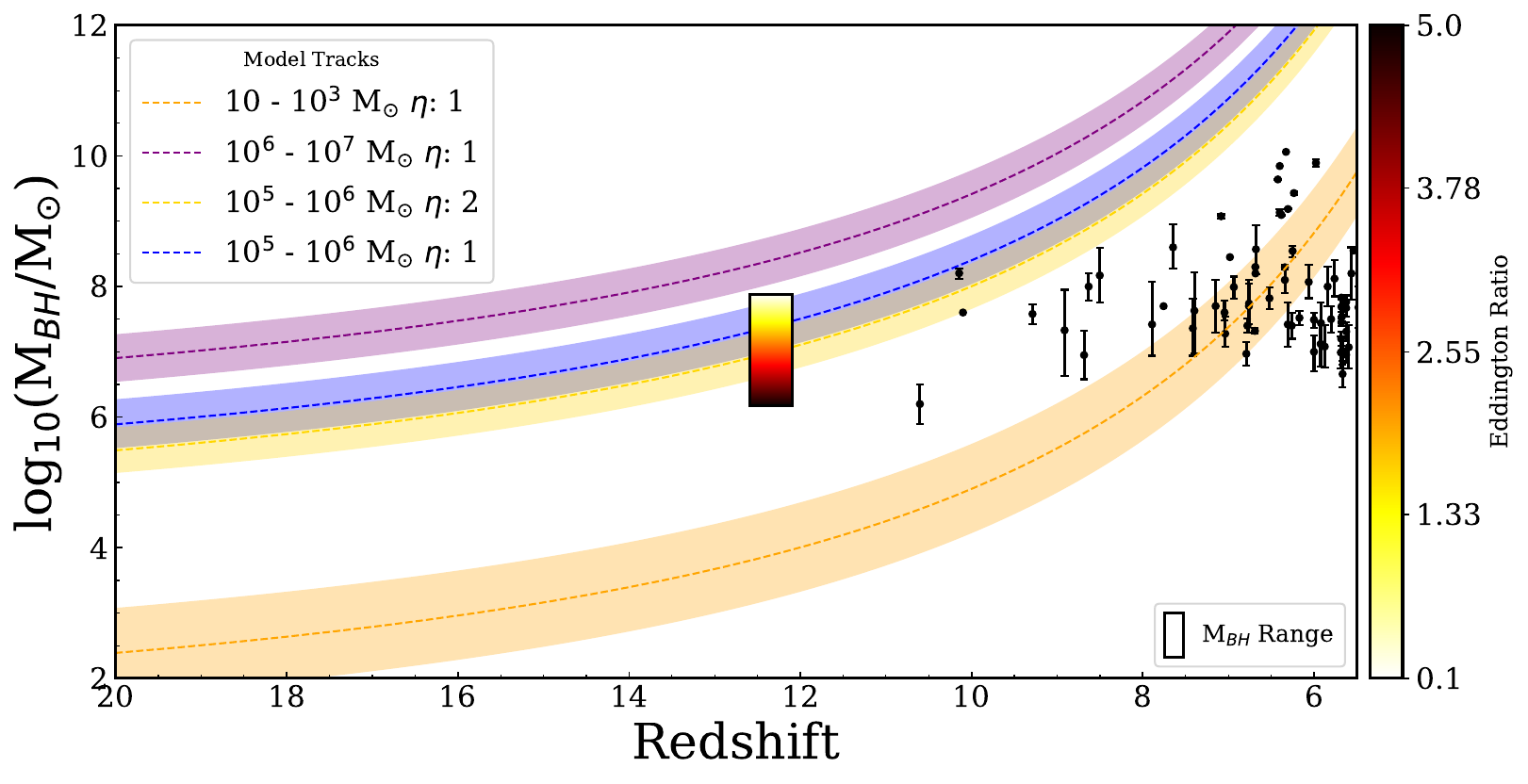}
    \caption{We show select SMBH seeding and growth scenarios, assuming a `light', stellar-remnant seed (100-300\,M$_{\odot}$), as well as a series of `heavy' seeds ($\gtrsim 10^5 {\rm \, M}_{\odot}$). In the scenarios shown here, the 'heavy' seed models are able to match the black hole mass of GHZ2, requiring extreme conditions for the black hole mass to be reached but are within the theoretical limits.  We also show for context other high redshift AGNs from the literature as the black points.  The rectangular gradient shows the range of black hole masses we infer from the best fitting model assuming a range of Eddington ratios 0.1 - 5, as shown by the colorbar.}
    \label{fig:bh_growth}
\end{figure}

Theoretical studies suggest that black hole seeds can reach higher masses at early times under extreme conditions. In highly-biased regions, corresponding to the tail in the Gaussian random field of density perturbations, stellar collapse can begin at $z >$ 20 \citep[e.g.,][]{Haemmer_bh_sim_2020}, providing a longer timeline for accretion. Furthermore, heavy seeds may reach 10$^8$ \Msun\ if rotation \citep{Shibata_2016, Dennison_2019} or dark collapse \citep{Mayer_Bonoli_2019} are considered, although typical heavy-seed masses are $\sim$10$^{5-6}$ \Msun \citep[e.g.,][]{Becerra_2018}. Invoking the self-annihilation of weakly interacting massive particles (WIMPs), ``dark stars" are also proposed as a channel for heavy seeds \citep[][]{Freese_2016}. Another non-standard pathway is linked to massive primordial black holes (PBHs), formed in the ultra-early universe \citep{Zhang_2025}, possibly able to explain the peculiar case of the `naked', extremely overmassive, SMBH in Abell~2744-QSO1, recently discovered by {\it JWST} \citep{Maiolino_PBH_2025}. 

Pursuing a different avenue towards accelerated early BH growth, multiple studies have explored super-Eddington accretion, including the conditions under which it can occur, and the duty cycles over which it may be sustained within realistic cosmological environments \citep[e.g.,][]{June_2023,Lupi_2024,Gordon_2025}. In any case, episodes of super-Eddington accretion may be required to explain the high-mass tail of the BH mass function at high redshifts, irrespective of the seeding model \citep[e.g.,][]{Jeon_MBH_Function_Sim_2025}.

To illustrate these possibilities, we construct an idealized BH growth model \citep{salpeter_bh_growth_1964,Haiman_2001}, under four sets of initial seeding and growth conditions: (1) a 10$^6$–10$^7$ \Msun\ heavy seed accreting at the Eddington limit at $z$ = 20, (2) a 10$^5$–10$^6$ \Msun\ heavy seed accreting at twice the Eddington limit at $z =$ 20, (3) a 10$^5$–10$^6$ \Msun\ heavy seed accreting at the Eddington limit at $z =$ 30, and (4) a stellar-remnant seed of 10–10$^{2.47}$ \Msun\ accreting at the Eddington limit at $z =$ 20. All models assume a radiative efficiency of 10\%.
Figure~\ref{fig:bh_growth} shows that, for select scenarios, this toy model can reproduce the inferred black hole mass of GHZ2 by $z$ = 12.34. The key factors required to reach this mass are: (i) a high initial seed mass ($>$10$^{5}$ \Msun), (ii) an early formation redshift (z $>$ 20), and (iii) super-Eddington accretion to enable efficient growth.

Linking observable properties to specific growth mechanisms—such as the direct collapse scenario, super-Eddington accretion, or early formation—would be particularly valuable. With ongoing theoretical studies and forthcoming observations of GHZ2, it will be possible to further constrain the formation and growth pathway of the AGN, which our analysis suggests to reside within this system. Given the current factor $\sim10$ uncertainty in the SMBH mass estimate, it remains an open question how severely GHZ2 may challenge our models of early SMBH formation. For the high-mass value, the system would be extremely overmassive, similar to Abell~2744-QSO1, but pushing this extreme configuration to even higher redshifts. For such a possible {\it in extremis} situation, all current SMBH seeding/growth models would be severely `stress tested'. However, on the positive side, we would have the opportunity to learn fundamental lessons about non-linear structure formation in the early universe \citep{MBK_2023}.

\subsection{Modeling Caveats}

While \BEAGLE-AGN represents a significant step forward, there are important caveats in its modeling approach. Notably, it fails to reproduce all the nitrogen lines in GHZ2, specifically \ion{N}{3}]$\lambda$1746, even in the best-fitting models. This limitation likely arises from the parameter space covered by the available grids. Addressing this issue will require careful consideration of nitrogen enrichment in high-redshift systems. Observations indicate that some high-redshift galaxies exhibit enhanced nitrogen abundances, and \citet{Cameron_2023} explores possible mechanisms, including nitrogen-loud quasars, Wolf-Rayet stars enriching the ISM via winds, and mergers of dense star clusters. Any of these processes could contribute to the enhanced nitrogen in GHZ2; however, further investigation is required to understand nitrogen enrichment under the extreme conditions near AGN at high redshift.

Nitrogen enrichment is not the only factor to consider; other ionization mechanisms may also contribute. For example, \citet{Flury_2024} highlight the role of shocks in powering nitrogen lines in enriched systems. Understanding the sources of nitrogen excitation is critical, as they can mimic features typically attributed to AGN activity. This emphasizes the need for models capable of capturing the full complexity of these systems and disentangling their various components. A detailed understanding of high-redshift extreme systems is therefore essential to accurately model nitrogen-to-oxygen enhancement and to identify the diverse physical processes responsible for their emission.

As described in Section 2.1, we exclude [\ion{Ne}{3}]$\lambda3869$ from the likelihood calculation. This decision is motivated by the documented inability of the current BEAGLE-AGN model grid to reproduce the observed range of [\ion{Ne}{3}]$\lambda$3869 emission in many high-redshift galaxies \citep{Silcock_2025}. Incorporating an observable that is not adequately represented by the model grid can bias the inferred physical parameters by forcing the fit to compensate through unrelated quantities such as metallicity or the AGN contribution. To verify that this choice does not affect our conclusions, we repeated the full analysis including [\ion{Ne}{3}]$\lambda3869$ and found that the posterior distributions of the primary galaxy and AGN parameters changed negligibly. Since the AGN properties in GHZ2 are primarily constrained by the high-ionization UV emission lines rather than [\ion{Ne}{3}]$\lambda3869$, we adopt the conservative approach of excluding this transition from the likelihood while avoiding the propagation of known model limitations into the inferred parameters. These limitations stem from the current photoionization models and correcting them lies beyond the scope of this work.

 A further caveat concerns the nature of the stellar populations at $z$=12.34, which remains highly uncertain. The dominant contributors could be Population III stars and/or very metal-poor, early Population II stars. In addition to uncertainty in the relative mix of these populations, their ionizing spectra and initial mass functions (IMFs) may differ substantially from those assumed in standard stellar population models, with direct consequences for the UV continuum and line emission. There is also significant uncertainty in the binary fraction at such early times and in the incidence of high-mass X-ray binaries, both of which can further modify the ionizing background.  A comprehensive assessment of these effects would require modeling a broad range of stellar population properties (e.g., Pop III vs. metal-poor Pop II, alternative IMFs, binary fractions) and their impact on the emergent spectrum, which is beyond the scope of this work.

Another caveat in our analysis concerns the treatment of gas density in the model grids. In \BEAGLE-AGN, the SF and AGN grids are generated separately with distinct density thresholds using the photoionization code \texttt{Cloudy} \citep{Ferland_Cloudy_2017}. High-redshift galaxies exist in environments that differ substantially from local systems, and the density of \HII\ regions directly affects level populations and the resulting emission lines. We tested only two nebular densities 10$^2$ and 10$^4$ cm$^{-3}$ with \BEAGLE-AGN and \texttt{Bagpipes}. However, it remains plausible that higher-density grids, modeled self-consistently including stars, \HII\ regions, and AGN, could improve these analyses. Exploring such extreme environments is beyond the scope of this work and is reserved for future studies.

We also note that while other higher density grids should be explored, another possibility is that of a multi-phase density distribution, where the system is composed of various densities all contributing together to make the final spectra that we see within GHZ2.  This can be especially true with the unique and different conditions in high-redshift galaxies. However, this is far beyond the scope of this paper as this would require new runs of photoionization models, taking careful consideration of the gas geometries and distribution, populating the gas abundance in each of the various phases, as well as careful consideration of the input ionization spectrum relative to the first density layer.

{Another implicit assumption in this analysis is that the emission-line measurements have sufficient signal-to-noise ratio (S/N) to constrain the physical properties of the combined AGN and star-forming regions. \citet{vidal_garcia_2024} tested \BEAGLE-AGN using synthetic spectra and found that an H$\beta$ S/N of $\sim10$ is required to robustly distinguish a NLR contribution when relying on the optical emission-line diagnostics alone. GHZ2 does not satisfy this criterion, as H$\beta$ is undetected. However, unlike the synthetic tests presented by \citet{vidal_garcia_2024}, the constraints on GHZ2 are not driven primarily by H$\beta$. Instead, the spectrum contains several high-S/N, high-ionization UV emission lines, including \CIV$\lambda$1548+1551 and \HeII$\lambda$1640, that are particularly sensitive to the hardness of the ionizing radiation field and the presence of an AGN-powered NLR. Consequently, although the H$\beta$ constraint is weak, the UV line diagnostics provide sufficient information to constrain the NLR properties and the AGN contribution within the \BEAGLE-AGN framework.

A final limitation of our modeling is that the AGN spectrum continuum is sub-dominant and does not contribute significantly to the overall model continuum emission due to the NLR assumption. While our current data show no evidence for a broad-line region, deeper higher-resolution spectra could reveal a broadened emission line such as broad carbon emission in \CIV, \CIII, or H$\alpha$ (a MIRI/MRS spectrum is incoming in Cycle 4 which may reveal broad H$\alpha$; PID 7078, PI: Mitsuhashi). Accounting for an AGN continuum would alter both the inferred AGN contribution to the emission lines and the derived galactic properties, as the observed photometry would be a combination of stellar and AGN continuum. For example, one property affected would be the stellar mass. Right now the stellar mass inferred is measured in the absence of an AGN continuum component, thus the stellar mass derived from \BEAGLE-AGN will be systematically overestimated. Since the continuum is modeled solely through stellar emission, the fit attributes all continuum light to stars, inflating the inferred stellar mass. Consequently, the derived black hole–to–stellar mass ratios should be interpreted as lower limits. 

This highlights how sensitive spectral modeling is to the inclusion of AGN components. Nevertheless, this modeling approach offers valuable insight into how different AGN components influence the spectrum. Despite these limitations in the current modeling framework, we have explored all feasible avenues with the available tools and consistently find significant AGN contributions in GHZ2 and draw cautious inferences about the properties of GHZ2 within these constraints.

\section{Summary}\label{sec:summary}

We present an analysis of GHZ2 to determine whether AGN activity can explain the strength of the high ionization UV lines with the presence of AGN activity. We use the \BEAGLE-AGN SED fitting code to disentangle the stellar, nebular, and AGN contribution to the overall SED of GHZ2 and summarize our findings below:

\begin{enumerate}
    \vspace{-2mm}
    \item The SF+AGN models are  statistically favored over a stellar only model via large differences in $\Delta \chi^2$ between the SF and SF+AGN models. Testing other SF scenarios with \texttt{Bagpipes} non-parametric modeling gave similar results.\\
    \vspace{-6mm}
    \item We find that the strength of the \CIV$\lambda$1548+1551 and \CIII]$\lambda$1908\ emission lines require an AGN contribution that constitutes 97$_{-2}^{+1}$\%, 64$_{-12}^{+23}$\% to the line flux. The AGN component also impacts other emission lines in a non-negligible way inferring values of  100$_{-1}^{+0}$\% for \HeII$\lambda$1640, 64$_{-5}^{+6}$\% for \OIII$\lambda$5007 and 45$_{-5}^{+6}$\% for H$\alpha$. While the optical line fluxes can be reproduced with SF models only, an AGN contribution is required for the  high ionization state UV emission lines.\\
    \vspace{-6mm}
    \item Making an assumption about the black hole accretion rate and using the accretion luminosity from \BEAGLE-AGN, we infer a black hole mass from the best fitting \BEAGLE-AGN model of log$_{10}$(M$_{BH}$/\Msun) =  6.89$^{+0.03}_{-0.03}$, 7.19$^{+0.03}_{-0.03}$, 7.89$^{+0.03}_{-0.03}$ assuming an Eddington ratio of 1, 0.5, and 0.1 respectively.  When testing the impact of the AGN power law slope on the derived black hole masses we find a systematic uncertainty of 0.6 dex.
    This $\sim$1 dex systematic uncertainty can be improved with future observations, such as PID 7078 (PI Mitsuhashi) which aims to detect a potential broad H$\alpha$ line.\\
    \vspace{-6mm}
    \item Using the derived black hole mass and the \BEAGLE-AGN stellar mass we compute a black-hole-to stellar mass ratio of: 0.03$^{+0.01}_{-0.01}$, 0.06$^{+0.02}_{-0.02}$, and  0.30$^{+0.09}_{-0.11}$for an Eddington ratio of 1, 0.5, and 0.1. 
    \item We find that GHZ2 likely supports a heavy seeding mechanism to achieve a black hole of its mass by redshift 12.34. To reach the derived black hole mass requires a high seeding mass (10$^{6}$ - 10$^{7}$ \Msun, $\eta$ = 1), seed formation at an earlier redshift (z $>$ 20, $\eta$ = 1), and/or super-Eddington accretion (10$^{5}$ - 10$^{6}$ \Msun, $\eta$ = 2). 
\end{enumerate}

The striking UV emission lines observed in GHZ2 make it a remarkable case for studying how stars and black holes evolved in the early universe \citep[e.g.,][]{Bromm_2011}. These lines could arise from either bursts of intense star formation or AGN activity, and previous studies \citep[see][]{Castellano_2024, Calabro_2024} have noted that GHZ2 sits in the “composite” region of several diagnostic diagrams—where both processes are likely at play. By explicitly modeling GHZ2 as a composite system using \BEAGLE-AGN, we find that including an AGN component is essential to reproduce the observed high-ionization features, pointing to a genuine contribution from an AGN alongside the stellar component.

This study is an initial step toward disentangling the AGN and star formation contributions in GHZ2, but it will not be the last.  The modeling carried out in this work relies on several assumptions—such as the treatment of the narrow-line region—that warrant further refinement with incorporating broad line fitting and AGN continuum emission. It is possible with other model assumption and model grids that the results could be altered and as such the quantitative AGN fraction and metallicity we infer should be viewed as model‑dependent estimates. The results highlight the need for more comprehensive theoretical frameworks that jointly incorporate AGN, stellar, and even shock-driven components when interpreting high-redshift galaxies. Only through such models can we fully capture the complexity of systems where multiple ionization sources shape the observed spectra and photometry.

 To complement the modeling, more data needs to be acquired to further validate and calibrate theoretical models. Especially in a system such as GHZ2, which reflects a unique and rare system capable of stress testing current limits.  Thus, more data and improved modeling will enable tighter constraints on key AGN properties, such as black hole mass—the fundamental observable for testing black hole growth models. As observations push the frontier to ever higher redshifts, refined analyses of these composite systems may help distinguish between different black hole seeding and growth scenarios, offering critical insight into the co-evolution of galaxies and their central black holes.

\software{Astropy \citep[][]{Astropy_2013, Astropy_2018, Astropy_2022}, 
Numpy \citep[][]{numpy_2020}
, Pandas \citep[][]{pandas_2010}, 
Scipy \citep[][]{Scipy_2020}, 
Matplotlib \citep[][]{matplotlib_2007}}

\begin{acknowledgments}
\section{Acknowledgments}
OCO thanks the University of Texas at Austin, and the NASA FINESST Fellowship for additional support.  OCO, and SLF acknowledge support from the NSF through NSF AAG award 1908817, and NASA through ADAP award 80NSSC22K0489 and the FINESST award 22-ASTRO22-0224.

MSS acknowledges support from an STFC PhD studentship (grant ST/V506709/1).

ECL acknowledges support of an STFC Webb Fellowship (ST/W001438/1)

AVG acknowledges support from the Spanish grant PID2022-138560NB-I00, funded by MCIN/AEI/10.13039/501100011033/FEDER, EU.

The authors acknowledge the Texas Advanced Computing Center (TACC) at The University of Texas at Austin for providing high-performance computing, visualization, and storage resources that have contributed to the research results reported in this article.

We also acknowledge that we did this work at an institution, the University of Texas at Austin, that sits on indigenous land. The Tonkawa live in central Texas and the Comanche and Apache move through this area. We pay respects to all the American Indians and indigenous peoples and communities that are a part of these lands and territories in Texas.  We are grateful to be able to live, work, collaborate, and learn on this piece of Turtle Island.
\end{acknowledgments}

\newpage
\bibliography{sample631}{}
\bibliographystyle{aasjournal}

\end{document}